\documentclass[dvipdfmx]{preprint}
\usepackage{geometry}
\usepackage{color}
\definecolor{note_fontcolor}{rgb}{0.578125, 0.542969, 0.542969}
\definecolor{shadecolor}{rgb}{0.941406, 0.941406, 0.847656}
\usepackage{verbatim}
\usepackage{framed}
\usepackage{mathrsfs}
\usepackage{amsmath}
\usepackage{amsthm}
\usepackage{amssymb}
\usepackage{stmaryrd}

\makeatletter

\newenvironment{lyxgreyedout}
  {\textcolor{note_fontcolor}\bgroup\ignorespaces}
  {\ignorespacesafterend\egroup}

\numberwithin{equation}{section}
\theoremstyle{plain}
\newtheorem{thm}{\protect\theoremname}[section]
\theoremstyle{plain}
\newtheorem{lem}[thm]{\protect\lemmaname}
\theoremstyle{plain}
\newtheorem{assumption}[thm]{\protect\assumptionname}
\theoremstyle{plain}
\newtheorem{cor}[thm]{\protect\corollaryname}
\theoremstyle{definition}
\newtheorem{defn}[thm]{\protect\definitionname}
\theoremstyle{plain}
\newtheorem{prop}[thm]{\protect\propositionname}
\theoremstyle{remark}
\newtheorem{rem}[thm]{\protect\remarkname}
\theoremstyle{definition}
\newtheorem{example}[thm]{\protect\examplename}
\theoremstyle{plain}
\newtheorem{conjecture}[thm]{\protect\conjecturename}
\theoremstyle{remark}
\newtheorem*{acknowledgement*}{\protect\acknowledgementname}

\usepackage{fancybox}
\usepackage{textcomp}
\usepackage[utf8]{inputenc}
\usepackage{lmodern}

\usepackage[destlabel,backref,dvipdfmx,bookmarksopenlevel=4,bookmarksnumbered]{hyperref}
\usepackage{color}
\usepackage{bbm}
\definecolor{magenta}{RGB}{30, 0, 50}
\definecolor{shadecolor}{rgb}{0.95,1,0.9}

\DeclareFontFamily{U}{mathx}{\hyphenchar\font45}
\DeclareFontShape{U}{mathx}{m}{n}{
      <5> <6> <7> <8> <9> <10>
      <10.95> <12> <14.4> <17.28> <20.74> <24.88>
      mathx10
      }{}
\DeclareSymbolFont{mathx}{U}{mathx}{m}{n}

\DeclareMathSymbol{\bigtimes}       {1}{mathx}{"91}



\makeatother

\providecommand{\acknowledgementname}{Acknowledgement}
\providecommand{\assumptionname}{Assumption}
\providecommand{\conjecturename}{Conjecture}
\providecommand{\corollaryname}{Corollary}
\providecommand{\definitionname}{Definition}
\providecommand{\examplename}{Example}
\providecommand{\lemmaname}{Lemma}
\providecommand{\propositionname}{Proposition}
\providecommand{\remarkname}{Remark}
\providecommand{\theoremname}{Theorem}

\begin{document}
\title{The Berezin--Simon quantization for K\"ahler manifolds and their
path integral representations}
\author{Hideyasu Yamashita}
\institute{Division of Liberal Arts and Sciences, Aichi-Gakuin University\\
\email{yamasita@dpc.aichi-gakuin.ac.jp}}

\maketitle

\renewenvironment{shaded}
  {\bgroup\ignorespaces}
  {\ignorespacesafterend\egroup}

\newenvironment{trivenv}
  {\bgroup\ignorespaces}
  {\ignorespacesafterend\egroup}

\newcommand{\displabel}[1]{\textcolor[rgb]{0.5, 0.5, 0.5}{\texttt{\textup{\tiny{}lab=#1}}}}

\newcommand{\hidable}[3]{#2}
\newcommand{\hidea}[1]{{#1}}
\newcommand{\hideb}[1]{{#1}}
\newcommand{\hidec}[1]{{#1}}
\newcommand{\hidep}[1]{{#1}}
\renewcommand{\hidec}[1]{}
\renewcommand{\hidep}[1]{}

\newcommand{\thlab}[1]{{\tt [#1]}}

\newenvironment{proofbar}
{\begin{leftbar}\noindent{\bf Proof.}}
{\noindent{\bf QED}\end{leftbar}}

\global\long\def\N{\mathbb{N}}%
\global\long\def\C{\mathbb{C}}%
\global\long\def\Z{\mathbb{Z}}%
 
\global\long\def\R{\mathbb{R}}%
 
\global\long\def\im{\mathrm{i}}%

\global\long\def\di{\partial}%
 
\global\long\def\d{{\rm d}}%

\global\long\def\ol#1{\overline{#1}}%
\global\long\def\ul#1{\underline{#1}}%
\global\long\def\ob#1{\overbrace{#1}}%

\global\long\def\ov#1{\overline{#1}}%

\global\long\def\then{\Rightarrow}%
 
\global\long\def\Then{\Longrightarrow}%

\global\long\def\al{\alpha}%
\global\long\def\de{\delta}%
 
\global\long\def\ep{\epsilon}%
 
\global\long\def\la{\lambda}%
 
\global\long\def\io{\iota}%
 
\global\long\def\th{\theta}%
\global\long\def\si{\sigma}%
 
\global\long\def\om{\omega}%

\global\long\def\De{\Delta}%
 
\global\long\def\Th{\Theta}%
 
\global\long\def\Om{\Omega}%

\global\long\def\brho{\boldsymbol{\rho}}%
\global\long\def\bDelta{\boldsymbol{\Delta}}%
 
\global\long\def\bmu{\boldsymbol{\mu}}%
 
\global\long\def\bchi{\boldsymbol{\chi}}%
 
\global\long\def\bPi{\boldsymbol{\Pi}}%
 
\global\long\def\bOm{\boldsymbol{\Omega}}%

\global\long\def\cA{\mathcal{A}}%
\global\long\def\cB{\mathcal{B}}%
 
\global\long\def\cC{\mathcal{C}}%
 
\global\long\def\cD{\mathcal{D}}%
\global\long\def\cE{\mathcal{E}}%
 
\global\long\def\cF{\mathcal{F}}%
 
\global\long\def\cG{{\cal G}}%
 
\global\long\def\cH{\mathcal{H}}%
 
\global\long\def\cI{\mathcal{I}}%
 
\global\long\def\cJ{\mathcal{J}}%
\global\long\def\cK{\mathcal{K}}%
 
\global\long\def\cL{\mathcal{L}}%
 
\global\long\def\cM{\mathcal{M}}%
 
\global\long\def\cN{\mathcal{N}}%
 
\global\long\def\cO{\mathcal{O}}%
 
\global\long\def\cP{\mathcal{P}}%
 
\global\long\def\cQ{\mathcal{Q}}%
 
\global\long\def\cR{\mathcal{R}}%
 
\global\long\def\cS{\mathcal{S}}%
 
\global\long\def\cT{\mathcal{T}}%
 
\global\long\def\cU{\mathcal{U}}%
 
\global\long\def\cV{\mathcal{V}}%
 
\global\long\def\cW{\mathcal{W}}%
\global\long\def\cX{\mathcal{X}}%
 
\global\long\def\cY{\mathcal{Y}}%
 
\global\long\def\cZ{\mathcal{Z}}%

\global\long\def\scA{\mathscr{A}}%
\global\long\def\scB{\mathscr{B}}%
\global\long\def\scC{\mathscr{C}}%
\global\long\def\scD{\mathscr{D}}%
 
\global\long\def\scE{\mathscr{E}}%
 
\global\long\def\scF{\mathscr{F}}%
 
\global\long\def\scG{\mathscr{G}}%
 
\global\long\def\scH{\mathscr{H}}%
 
\global\long\def\scI{\mathscr{I}}%
 
\global\long\def\scJ{\mathscr{J}}%
 
\global\long\def\scK{\mathscr{K}}%
 
\global\long\def\scL{\mathscr{L}}%
 
\global\long\def\scM{\mathscr{M}}%
 
\global\long\def\scN{\mathscr{N}}%
 
\global\long\def\scO{\mathscr{O}}%
 
\global\long\def\scP{\mathscr{P}}%
 
\global\long\def\scR{\mathscr{R}}%
\global\long\def\scS{\mathscr{S}}%
 
\global\long\def\scT{\mathscr{T}}%
 
\global\long\def\scU{\mathscr{U}}%
 
\global\long\def\scZ{\mathscr{Z}}%

\global\long\def\bbA{\mathbb{A}}%
 
\global\long\def\bbB{\mathbb{B}}%
 
\global\long\def\bbD{\mathbb{D}}%
 
\global\long\def\bbF{\mathbb{F}}%
 
\global\long\def\bbG{\mathbb{G}}%
 
\global\long\def\bbI{\mathbb{I}}%
 
\global\long\def\bbK{\mathbb{K}}%
 
\global\long\def\bbL{\mathbb{L}}%
 
\global\long\def\bbM{\mathbb{M}}%
 
\global\long\def\bbP{\mathbb{P}}%
 
\global\long\def\bbQ{\mathbb{Q}}%
 
\global\long\def\bbT{\mathbb{T}}%
 
\global\long\def\bbU{\mathbb{U}}%
 
\global\long\def\bbX{\mathbb{X}}%
 
\global\long\def\bbY{\mathbb{Y}}%
\global\long\def\bbW{\mathbb{W}}%

\global\long\def\bbOne{1\kern-0.7ex  1}%

\renewcommand{\bbOne}{\mathbbm{1}}

\global\long\def\bB{\mathbf{B}}%
 
\global\long\def\bG{\mathbf{G}}%
 
\global\long\def\bH{\mathbf{H}}%
 
\global\long\def\bM{\mathbf{M}}%
 
\global\long\def\bS{\boldsymbol{S}}%
 
\global\long\def\bT{\mathbf{T}}%
 
\global\long\def\bX{\mathbf{X}}%
\global\long\def\bY{\mathbf{Y}}%
\global\long\def\bW{\mathbf{W}}%
 
\global\long\def\boT{\boldsymbol{T}}%

\global\long\def\fraka{\mathfrak{a}}%
 
\global\long\def\frakb{\mathfrak{b}}%
 
\global\long\def\frakc{\mathfrak{c}}%
 
\global\long\def\frake{\mathfrak{e}}%
 
\global\long\def\frakf{\mathfrak{f}}%
 
\global\long\def\fg{\mathfrak{g}}%
 
\global\long\def\frakh{\mathfrak{h}}%
 
\global\long\def\fraki{\mathfrak{i}}%
 
\global\long\def\frakk{\mathfrak{k}}%
 
\global\long\def\frakl{\mathfrak{l}}%
 
\global\long\def\frakm{\mathfrak{m}}%
 
\global\long\def\frakn{\mathfrak{n}}%
 
\global\long\def\frako{\mathfrak{o}}%
 
\global\long\def\frakp{\mathfrak{p}}%
 
\global\long\def\frakq{\mathfrak{q}}%
 
\global\long\def\frakr{\mathfrak{r}}%
 
\global\long\def\fs{\mathfrak{s}}%
 
\global\long\def\frakt{\mathfrak{t}}%
 
\global\long\def\fraku{\mathfrak{u}}%

\global\long\def\fA{\mathfrak{A}}%
 
\global\long\def\fB{\mathfrak{B}}%
 
\global\long\def\fC{\mathfrak{C}}%
 
\global\long\def\fD{\mathfrak{D}}%
 
\global\long\def\fF{\mathfrak{F}}%
 
\global\long\def\fG{\mathfrak{G}}%
 
\global\long\def\fK{\mathfrak{K}}%
 
\global\long\def\fL{\mathfrak{L}}%
 
\global\long\def\fM{\mathfrak{M}}%
 
\global\long\def\fP{\mathfrak{P}}%
 
\global\long\def\fR{\mathfrak{R}}%
 
\global\long\def\fT{\mathfrak{T}}%
 
\global\long\def\fU{\mathfrak{U}}%
 
\global\long\def\fX{\mathfrak{X}}%

\global\long\def\hM{\hat{M}}%

\global\long\def\rM{\mathrm{M}}%
\global\long\def\prj{\mathfrak{P}}%

{} 
\global\long\def\sy#1{{\color{blue}#1}}%

\global\long\def\magenta#1{{\color{magenta}#1}}%

\global\long\def\symb#1{#1}%

{} %

\global\long\def\emhrb#1{\text{{\color{red}\huge{\bf #1}}}}%

\newcommand{\symbi}[1]{\index{$ #1$}{\color{red}#1}} 

{} 
\global\long\def\SYM#1#2{#1}%

\renewcommand{\SYM}[2]{\symb{#1}}

\newcommand{\usuji}{\color[rgb]{0.7,0.4,0.4}} \newcommand{\usu}{\color[rgb]{0.5,0.2,0.1}}
\newenvironment{Usuji} {\begin{trivlist}   \item \usuji }  {\end{trivlist}}
\newenvironment{Usu} {\begin{trivlist}   \item \usu }  {\end{trivlist}} 

\newcommand{\term}[1]{\textcolor[rgb]{0, 0, 1}{\bf #1}}
\newcommand{\termi}[1]{{\bf #1}}

\global\long\def\rG{\mathrm{G}}%
 
\global\long\def\rT{\mathrm{T}}%
 
\global\long\def\rH{\mathrm{H}}%
 
\global\long\def\rU{\mathrm{U}}%

\global\long\def\supp{{\rm supp}}%
\global\long\def\dom{\mathrm{dom}}%
\global\long\def\ran{\mathrm{ran}}%
 
\global\long\def\leng{\text{{\rm leng}}}%
 
\global\long\def\diam{\text{{\rm diam}}}%
 
\global\long\def\Leb{\text{{\rm Leb}}}%
 
\global\long\def\meas{\text{{\rm meas}}}%
\global\long\def\sgn{{\rm sgn}}%
 
\global\long\def\Tr{{\rm Tr}}%
 
\global\long\def\tr{\mathrm{tr}}%
 
\global\long\def\spec{{\rm spec}}%
 
\global\long\def\Ker{{\rm Ker}}%
 
\global\long\def\Lip{{\rm Lip}}%
 
\global\long\def\Id{{\rm Id}}%
 
\global\long\def\id{{\rm id}}%

\global\long\def\ex{{\rm ex}}%
 
\global\long\def\Pow{\mathsf{P}}%
 
\global\long\def\Hom{\mathrm{Hom}}%
 
\global\long\def\grad{\mathrm{grad}}%
 
\global\long\def\End{{\rm End}}%
 
\global\long\def\Aut{{\rm Aut}}%

\newcommand{\slim}{\mathop{\mbox{\rm s-lim}}} %

\newcommand{\wlim}{\mathop{\mbox{\rm w-lim}}}

\newcommand{\limsub}{\mathop{\mbox{\rm lim-sub}}}

\global\long\def\bboxplus{\boxplus}%

\renewcommand{\bboxplus}{\mathop{\raisebox{-0.8ex}{\text{\begin{trivenv}\LARGE{}$\boxplus$\end{trivenv}}}}}

\global\long\def\shuff{\sqcup\kern-0.3ex  \sqcup}%

\renewcommand{\shuff}{\shuffle}

\global\long\def\upha{\upharpoonright}%

\global\long\def\ket#1{|#1\rangle}%
 
\global\long\def\bra#1{\langle#1|}%

{} 
\global\long\def\lll{\vert\kern-0.25ex  \vert\kern-0.25ex  \vert}%
 \renewcommand{\lll}{{\vert\kern-0.25ex  \vert\kern-0.25ex  \vert}}

\global\long\def\biglll{\big\vert\kern-0.25ex  \big\vert\kern-0.25ex  \big\vert\kern-0.25ex  }%
 
\global\long\def\Biglll{\Big\vert\kern-0.25ex  \Big\vert\kern-0.25ex  \Big\vert}%

\newcommand{\iiia}[1]{{\left\vert\kern-0.25ex\left\vert\kern-0.25ex\left\vert #1
  \right\vert\kern-0.25ex\right\vert\kern-0.25ex\right\vert}}

\global\long\def\iii#1{\iiia{#1}}%

\global\long\def\Upa{\Uparrow}%
 
\global\long\def\Nor{\Uparrow}%

\newcommand{\vertt}{\kern-0.6ex\vert}
\renewcommand{\Nor}{[\kern-0.16ex ]}

\global\long\def\Prob{\mathbb{P}}%
\global\long\def\Var{\mathrm{Var}}%
\global\long\def\Cov{\mathrm{Cov}}%
\global\long\def\Ex{\mathbb{E}}%
{} 
\global\long\def\Ae{{\rm a.e.}}%
 
\global\long\def\samples{\bOm}%

\global\long\def\var{\textrm{{\rm var}}}%
\global\long\def\Hol{\text{{\rm Höl}}}%
 
\global\long\def\hvar{\textrm{{\rm -var}}}%
\global\long\def\hHol{\text{{\rm -Höl}}}%

\global\long\def\pvar{p\textrm{{\rm -var}}}%
\global\long\def\pHol{1/p\text{{\rm -Höl}}}%
\global\long\def\frakt{\mathfrak{t}}%

\global\long\def\var{\textrm{{\rm var}}}%
\global\long\def\Hol{\text{{\rm Höl}}}%
 
\global\long\def\hvar{\textrm{{\rm -var}}}%
\global\long\def\hHol{\text{{\rm -Höl}}}%

\global\long\def\rpvar{\mathfrak{p}}%
 
\global\long\def\rpHol{\mathfrak{h}}%

\global\long\def\bOne{{\bf 1}}%

\global\long\def\Disk{\mathbb{D}^{2}}%
\global\long\def\hcG{\hat{\mathcal{G}}}%
\global\long\def\sfC{\mathsf{C}}%

{} 
\global\long\def\crv{\mathfrak{c}}%
\global\long\def\Crv{\mathfrak{C}}%
 
\global\long\def\gE{\mathrm{e}}%
 
\global\long\def\Rot{{\rm Rot}}%

\global\long\def\bbm#1{\mathbbm{#1}}%

\global\long\def\Mat{{\rm Mat}}%

\global\long\def\cbo{{\bf c}}%
 
\global\long\def\reg{{\rm reg}}%

\global\long\def\decoFor{\mathsf{DF}}%
 
\global\long\def\DF{\mathsf{DF}}%

{} 
\global\long\def\modsp{\scT}%
\global\long\def\regStr{\boldsymbol{T}}%

\global\long\def\smoothfuncs{\scC}%
 
\global\long\def\jj{\mathtt{j}}%
 
\global\long\def\scriptj{\mathtt{j}}%

\global\long\def\newNode{\circledast}%
 
\global\long\def\scriptf{\mathtt{f}}%
 
\global\long\def\scripth{\mathtt{h}}%

\global\long\def\p{\mathbf{p}}%
 
\global\long\def\q{\mathbf{q}}%
 
\global\long\def\bA{\mathbf{A}}%
 
\global\long\def\x{\mathbf{x}}%

\global\long\def\div{\mathrm{div}}%
 
\global\long\def\be{\beta}%
 
\global\long\def\La{\Lambda}%
 
\global\long\def\Ga{\Gamma}%

\global\long\def\wick#1{:\!#1\!:}%
 
\global\long\def\dag{\dagger}%

\global\long\def\braket#1{\langle#1\rangle}%
 
\global\long\def\ka{\kappa}%
 
\global\long\def\z{\mathbf{z}}%

\global\long\def\tI{t_{\mathrm{I}}}%
 
\global\long\def\tF{t_{\mathrm{F}}}%

\global\long\def\ActionIntegral{\mathrm{AI}}%

\global\long\def\Hc{h}%

\global\long\def\coherents{\mathbf{c}}%
 
\global\long\def\ssW{\mathsf{W}}%

\global\long\def\Ten{\bullet}%
{} %

\global\long\def\TT{\intercal}%
 \renewcommand{\TT}{\mathsf{T}}

\global\long\def\trit{\vartriangle\!\! t}%

\global\long\def\Killing{{\rm \boldsymbol{\kappa}}}%
 
\global\long\def\spec{{\rm spec}}%

\global\long\def\nnn{\mathfrak{k}}%

\global\long\def\Ad{{\rm Ad}}%

\global\long\def\Gxz{G\cdot x_{0}}%
 
\global\long\def\lbundle{\scL_{\lambda}}%
 
\global\long\def\Hilb{\cH_{\lambda}}%
 
\global\long\def\Image{{\rm Im}}%
 
\global\long\def\tautlog{{\rm taut}}%
 
\global\long\def\sphere{\mathbb{S}}%
 
\global\long\def\Proj{\mathbf{pj}}%
 
\global\long\def\telem{\mathbf{t}}%
 
\global\long\def\sectio{\mathbf{sect}}%

\global\long\def\hwvec{\mathbf{v}_{\lambda}}%
{} 
\global\long\def\Hwvproj{{\bf E}_{\lambda}}%

\global\long\def\lwvec{{\bf w}_{\lambda}}%

\global\long\def\lbundle{\mathscr{L}_{\lambda}}%
 
\global\long\def\Gxz{G\cdot x_{0}}%

\global\long\def\Rtrans{{\rm Rt}}%
 
\global\long\def\Ltrans{{\rm Lt}}%

\global\long\def\Rght{{\rm R}}%
 
\global\long\def\Lft{{\rm L}}%
 
\global\long\def\Casi{{\bf c}}%
 
\global\long\def\Rtrans{\mathscr{T_{\Rght}}}%
 
\global\long\def\Ltrans{\mathscr{T_{{\rm \Lft}}}}%
 
\global\long\def\Maurer{\mathscr{M}}%
 
\global\long\def\du{\underline{{\rm d}}}%
 
\global\long\def\bphi{\boldsymbol{\varphi}}%

{} 
\global\long\def\ss{{\bf r}}%
 
\global\long\def\infspec{{\bf c}_{\lambda}}%

\global\long\def\Roots{\boldsymbol{R}}%
 
\global\long\def\Sn{{\rm sig}}%
 
\global\long\def\lift{{\rm lift}}%
 
\global\long\def\bpi{\boldsymbol{\pi}}%

\global\long\def\Tensor{\boldsymbol{T}}%
 
\global\long\def\bOmega{\boldsymbol{\Omega}}%
 
\global\long\def\VECSP{\mathbb{V}}%
 
\global\long\def\projection{{\rm pr}}%
 
\global\long\def\dissect{\cD}%
 
\global\long\def\bUpsilin{\boldsymbol{\Upsilon}}%
 
\global\long\def\GEOMR{\mathbf{GR}}%

\global\long\def\fextend#1{\hat{#1}}%
 
\global\long\def\eigenval{\varepsilon}%

\global\long\def\vecalt{\check{\chi}}%
 
\global\long\def\Tanalt{\check{\cA}}%
 
\global\long\def\taut{{\rm taut}}%
 
\global\long\def\tautsection{^{!}\Gamma_{{\rm taut}}^{\infty}}%
\global\long\def\dtautsection{^{!}\Gamma_{{\rm taut}^{*}}^{\infty}}%
 
\global\long\def\holodtautsection{^{!}\Gamma_{{\rm taut}^{*}}^{{\rm hol}}}%
  
\global\long\def\Proj{{\bf pr}}%

\global\long\def\dequantize{\cR}%
 
\global\long\def\FCP{{\rm Coh}}%
 
\global\long\def\Manifold{\grave{{\bf M}}}%
\renewcommand{\Manifold}{{\bf M}}

\global\long\def\Ran{{\rm Ran}}%
 
\global\long\def\qproj{\mathsf{P}}%

\global\long\def\Hc{\grave{H}_{{\rm c}}}%

\global\long\def\dequantize{\cR}%
 
\global\long\def\FCP{{\rm Coh}}%
\global\long\def\Coh{{\rm Coh}}%

\global\long\def\Manifold{{\bf M}}%

\global\long\def\EP{E_{\bbP}}%

\global\long\def\Ran{{\rm Ran}}%
 
\global\long\def\qproj{\mathsf{P}}%

\global\long\def\Taut{{\rm Taut}}%
\global\long\def\bbS{\mathbb{S}}%
\global\long\def\vol{{\rm vol}}%

\global\long\def\Kpot{{\bf h}}%

\global\long\def\ela{e_{\io}}%

\global\long\def\INDEX{\cI}%

\global\long\def\CC{\Gamma}%

\global\long\def\INT{\scI}%

\global\long\def\scV{\mathscr{V}}%
 
\global\long\def\loc{{\rm loc}}%

\global\long\def\path{{\bf x}}%
\global\long\def\Path{{\bf X}}%

\global\long\def\no{\boldsymbol{\nu}}%

\global\long\def\dil{{\d^{\prime}}}%
\global\long\def\dill{{\d^{\prime\prime}}}%
\global\long\def\forms{{\bf A}}%
\global\long\def\ptrans{//}%

\global\long\def\Harmonic{{\rm Hm}}%

\def\mininabla{\text{\tiny $\nabla$}}
\def\Denabla{{\Delta_{\mininabla}}}

\global\long\def\foreignlanguage#1#2{#2}

\newcommand{\HS}{{\bf HS}}

\global\long\def\inv{-1}%
\global\long\def\ppa{{\bf t}}%
\global\long\def\zpath{{\bf z}}%
\global\long\def\ebold{{\bf e}}%

\begin{abstract}
The Berezin--Simon (BS) quantization is a rigorous version of the
``operator formalism'' of quantization procedure. The goal of the
paper is to present a rigorous real-time (not imaginary-time) path-integral
formalism corresponding to the BS operator formalism of quantization;
Here we consider the classical systems whose phase space $M$ is a
(possibly non-compact) K\"ahler manifold which satisfies some conditions,
with a Hamiltonian $H:M\to\R$. For technical reasons, we consider
only the cases where $H$ is smooth and bounded. We use G\"uneysu's
extended version of the Feynman--Kac theorem to formulate the path-integral
formula.
\end{abstract}

\section{Introduction}

In this paper, we set the ``classical phase space'' $\Manifold$
to be a (possibly non-compact) K\"ahlerian manifold which is a submanifold
of a (possibly infinite-dimensional) complex projective space $\bbP\cH$,
where $\cH$ is a complex Hilbert space. The phase space $\Manifold$
admits a quantization procedure, which we called the \emph{Glauber-Sudarshan-type
quantization} in \cite{Yam18}; However instead we will call it the
\emph{Berezin--Simon} (BS) quantization in this paper, since we follow
the formulation of the quantization procedure given in Simon \cite{Sim80},
which is based on Berezin's works; See \cite{Sim80} and references
therein. A BS quantization is an ``operator formalism'' of quantization
procedure. The goal of the paper is to present a path-integral representation
of the BS quantization; Roughly, we present a rigorous path-integral
formalism corresponding to the BS operator formalism of quantization.

The previous paper \cite{Yam18} had a similar goal, but it was very
restricted in that we confined ourselves to the cases where the phase
space $\Manifold$ is a compact homogeneous space, which is a submanifold
of a projective space $\bbP\cH$ where $\cH$ is a finite-dimensional
Hilbert space. Thus this paper will be viewed as a considerable extension
of \cite{Yam18}. 

Our main mathematical tool for the path-integral formulation is the
\emph{Feynman-Kac formula }on a vector bundle over a Riemannian manifold
given by \cite{Gun10}, together with the \emph{Bochner-Kodaira-Nakano
identity} for K\"ahlerian manifolds. A Feynman-Kac formula itself
is seen as a mathematical justification of the \emph{imaginary-time}
path integral method in quantum physics, but we devise a method to
use it for \emph{real-time} path integrals.

Roughly speaking, this paper is situated in the following context
of past rigorous studies on path integrals. 

{} Feynman's original idea \cite{FH65} is to represent the time evolution
of a quantum system, as well as the expectation values of observables
in it, by an integral on the space of paths on the \emph{configuration
space} of the system. As is well known, if we consider the ``imaginary
time'' evolution instead of real time evolution, so-called the \emph{Wick
rotation}, a large part of the idea can be made rigorous by the Feynman-{}-Kac
theorem and its generalizations, and this ``imaginary time $+$ Feynman--Kac''
approach is the most successful one. However, note that in the imaginary-time
approaches, it is difficult to deal with time-dependent Hamiltonians,
as well as non-unitary time evolutions occurring in open systems.
This implies that it is hard to apply the imaginary-time methods to
e.g.~the theories of quantum information/probability, where time-dependent
Hamiltonians and non-unitary time evolutions (e.g. decoherences) frequently
occur.

On the other hand, the notion on configuration-space path integrals
are believed to be derived from more general notion of \emph{phase-space}
path integrals. In some sense, the latter ones may be more fundamental
if we consider a path integral as a procedure of quantization of a
classical system; The main stream of the rigorous studies of quantization
(e.g. the theories of geometric/deformation quantization) are formulated
on phase spaces. Unlike imaginary-time configuration-space path integrals,
little is known about the rigorous justification of general phase-space
path integrals (in real or imaginary time).

However in 1985, Daubechies and Klauder \cite{DK85} gave an important
rigorous result on \emph{coherent-state path integrals}, which can
be viewed as a sort of phase-space path integral formula, representing
real-time evolution for some class of Hamiltonians, in terms of Brownian
motions and stochastic integrals. Yamashita \cite{Yam11} studied
phase-space path integrals in a similar idea but for other class of
Hamiltonians, and with an emphasis on geometric meaning of them. In
these results, mainly the phase spaces are assumed to be flat, i.e.,
$\Manifold\cong\R^{2n}\cong\C^{n}$. %
Yamashita \cite{Yam18} is seen as an attempt to apply such methods
on some sort of (non-flat) compact phase spaces, which arise in irreducible
unitary representation of semisimple Lie groups, e.g., ${\rm SU}(n)$,
${\rm SO}(n)$, ${\rm Sp}(n)$, etc. Then we are given a question
whether these works \cite{Yam11,Yam18} can be unified and extended
for more general phase spaces, or not. This paper is intended to be
an affirmative answer to this question. 

However note that we consider only the \emph{bounded} Hamiltonians
in this paper, hence not all of the results of \cite{DK85} is contained
in our result. Since this boundedness assumption is quite unsatisfactory
for applications to realistic physical systems, we are required to
loosen this assumption, but the treatment of general unbounded Hamiltonians
appears to be extremely difficult. A hopeful approach will be to examine
some moderate assumptions such as that the classical Hamiltonian $H(x)$
is bounded from below and increases as $H(x)\sim|x|^{2}$; Another
hopeful approach will be to consider a ``solvable'' (or ``algebraically
tractable'') set of Hamiltonians which are generators of (representations
of) a Lie group, e.g.~symplectic groups, Poincare groups, etc. The
latter approach will be related to the construction of unitary representations
of a Lie group in terms of orbit method/geometric quantization \cite{Kir04,Woo92}.

\section{Projective representations of BS quantizations}

\label{sec:BSquantization}

\subsection{BS quantization}

\begin{lyxgreyedout}
\hidec{

$\bbP\cH$ is naturally given a complex manifold structure when $\dim\cH<\infty$,
so that $\Proj$ is a holomorphic map. Even when $\dim\cH=\infty$,
$\bbP\cH$ is viewed as an infinite-dimensional Hilbert manifold with
the complex structure determined by $\Proj$. Furthermore, $\bbP\cH$
is given a natural Riemannian metric; $\bbP\cH$ is a metric space
with the distance function 
\begin{align*}
 & d(x,y):=2^{-1/2}\left\Vert x-y\right\Vert _{\HS}\quad(x,y\in\bbP\cH),\quad\text{or equivalently,}\\
 & d\left(\Proj(v),\Proj(u)\right):=\sqrt{1-\left|\left\langle v|u\right\rangle \right|^{2}},\quad u,v\in\cH,\ \left\Vert u\right\Vert =\left\Vert v\right\Vert =1.
\end{align*}
 where $\left\Vert x\right\Vert _{\HS}:=\left(\Tr x^{*}x\right)^{1/2}$
is the Hilbert--Schmidt norm. For $u,v\in\cH$ with $\left\Vert v\right\Vert =1,\left\langle v|u\right\rangle =0$,
and a smooth function $f$ on $\bbP\cH$, let
\[
D_{u}f(v):=\frac{\d}{\d t}\Big|_{t=0}f(\Proj(v+tu)).
\]
Then $d$ determines the norm on the tangent space $T_{\Proj(v)}\bbP\cH$
by

\[
\left\Vert D_{u}\right\Vert :=\lim_{t\searrow0}t^{-1}d\left(\Proj(v+tu),\ \Proj(v)\right)
\]
and the Riemannian metric
\[
g\left(D_{u_{1}},D_{u_{2}}\right):=\frac{1}{4}\left(\left\Vert D_{u_{1}}+D_{u_{2}}\right\Vert ^{2}-\left\Vert D_{u_{1}}-D_{u_{2}}\right\Vert ^{2}\right),\qquad u_{1},u_{2}\in\left\{ v\right\} ^{\perp}.
\]
Note that if $\Id:\bbP\cH\to\bbP\cH$ is the identity map, then the
Riemannian metric $g$ is expressed as 
\[
g(D_{u_{1}},D_{u_{2}})=\frac{1}{2}\left\langle D_{u_{1}}\Id|D_{u_{2}}\Id\right\rangle _{\HS}=\Re\left\langle u_{1}|u_{2}\right\rangle ,\qquad u_{1},u_{2}\in\left\{ v\right\} ^{\perp}.
\]
where $\left\langle X|Y\right\rangle _{\HS}:=\Tr X^{*}Y$. Hence $g$
is compatible with the complex structure, i.e., $g\left(D_{\im u_{1}},D_{\im u_{2}}\right)=g\left(D_{u_{1}},D_{u_{2}}\right)$
for $u_{1},u_{2}\in\left\{ v\right\} ^{\perp}$.

Define the 2-form $\omega$ by
\[
\omega(D_{u_{1}},D_{u_{2}}):=g(D_{\im u_{1}},D_{u_{2}})=\Re\left\langle \im u_{1}|u_{2}\right\rangle =\Im\left\langle u_{1}|u_{2}\right\rangle .
\]
Then we see $(\bbP\cH,g,\omega)$ is a (possibly infinite-dimensional)
K\"ahler manifold with the K\"ahler form $\omega$.

}%
\end{lyxgreyedout}

Let $\cH$ be a complex Hilbert space, and $\SYM{\bbP\cH}{PH}$ denote
the set of orthogonal projections onto one-dimensional subspaces of
$\cH$, that is,
\[
\SYM{\bbP\cH}{PH}:=\left\{ \ket v\bra v:\ v\in\cH,\ \|v\|=1\right\} .
\]
Let $\cH^{\times}:=\cH\setminus\{0\}$, and define $\Proj:\cH^{\times}\to\bbP\cH$
to be the map from $v\in\cH^{\times}$ to the orthogonal projection
onto $\C v$, i.e.,
\[
\SYM{\Proj(v)}{pr}:=\frac{\ket v\bra v}{\left\Vert v\right\Vert ^{2}},\qquad v\in\cH^{\times}.
\]

Let $\Manifold$ be a subset of $\bbP\cH$ with the measure $\mu$.
The measure space $(\Manifold,\mu)$ is called a \termi{family of coherent states}
on $\cH$ if 
\[
\int_{\Manifold}\p\,\d\mu(\p)=I.
\]
Let $\SYM{\Coh(\cH)}{Coh}$ be the set of families of coherent states
on $\cH$. For a function $f:\bbP\cH\to\C$, let
\[
\SYM{\cQ(f)}{Q(f)}:=\int_{\Manifold}f(\p)\p\,\d\mu(\p),
\]
if the integral exists. In this paper, we call the operation $f\mapsto\cQ(f)$
the \termi{BS quantization} on $(\Manifold,\mu)$.

Let

\[
\SYM{\Proj^{-1}(\Manifold)}{pr-1(M)}:=\left\{ v\in\cH^{\times}|\Proj(v)\in\Manifold\right\} =\bigcup_{\p\in\Manifold}\ran(\p)\setminus\{0\}
\]
\[
\SYM{\bbS(\cH)}{S(H)}:=\left\{ v\in\cH|\left\Vert v\right\Vert =1\right\} ,\quad\bbS(\Manifold):=\bbS(\cH)\cap\Proj^{-1}(\Manifold)
\]
The projection $\Proj|_{\bbS(\Manifold)}:\bbS(\Manifold)\to\Manifold$
defines a $U(1)$-bundle over $\Manifold$. (Precisely, the term ``bundle''
should be used only when $\Manifold$ is a (smooth) manifold.) Let
$\mu_{\bbS}$ be the $U(1)$-invariant measure on $\bbS(\Manifold)$
determined by 
\[
\mu(E)=\mu_{\bbS}(\Proj_{\bbS}^{-1}(E)),\qquad\forall E\subset\Manifold,\ \text{measurable}
\]
where $\Proj_{\bbS}:=\Proj|_{\bbS(\Manifold)}$. For measurable functions
$f_{1},f_{2}:\bbS(\Manifold)\to\C$ (or $f_{1},f_{2}:\Proj^{-1}(\Manifold)\to\C$),
let
\[
\left\langle f_{1}|f_{2}\right\rangle _{\bbS}:=\int_{\bbS(\Manifold)}\ol{f_{1}(v)}f_{2}(v)\d\mu_{\bbS}(v),
\]
if the integral exists. 

Let $\SYM{\cH^{*}}{H*}$ denote the dual of $\cH$. For any $u\in\cH$,
define $\SYM{u^{*}}{u*}\in\cH^{*}$ by 
\[
u^{*}(v):=\left\langle u|v\right\rangle ,\qquad v\in\cH.
\]
We denote $u^{*}|_{\bbS(\Manifold)}$ simply by $u^{*}$. Then we
find %
that 
\[
\left\langle u_{1}^{*}|u_{2}^{*}\right\rangle _{\bbS}=\left\langle u_{2}|u_{1}\right\rangle ,\qquad\forall u_{1},u_{2}\in\cH.
\]

The inner product $\left\langle \cdot|\cdot\right\rangle _{\bbS}$
defines the Hilbert space $L^{2}(\bbS(\Manifold))\equiv L^{2}(\bbS(\Manifold),\mu_{\bbS})$. 

For $\ell\in\Z$, let
\begin{align*}
\SYM{\CC_{\ell}(\cH^{\times})}{Gammal(H)} & :=\left\{ f:\cH^{\times}\to\C\,:\,f(\lambda v)=\lambda^{n}f(v),\ \ \forall\lambda\in\C^{\times},\ v\in\cH^{\times}\right\} ,\\
\SYM{\CC_{\ell,\Manifold}}{GammalM}\equiv\CC_{\ell}(\Proj^{-1}(\Manifold)) & :=\left\{ f:\Proj^{-1}(\Manifold)\to\C\,:\,f(\lambda v)=\lambda^{\ell}f(v),\ \ \forall\lambda\in\C^{\times},\ v\in\Proj^{-1}(\Manifold)\right\} ,\\
\SYM{\CC_{\ell,\Manifold}^{L^{p}}}{GammaL2} & :=\left\{ f\in\CC_{\ell,\Manifold}:\,f|_{\bbS(\Manifold)}\in L^{p}(\bbS(\Manifold),\mu_{\bbS})|\right\} ,\qquad1\le p\le\infty.
\end{align*}
For each $\ell\in\Z$, $\CC_{\ell,\Manifold}^{L^{2}}$ is a Hilbert
space with the inner product $\left\langle \cdot|\cdot\right\rangle _{\bbS}$,
and naturally viewed as a closed subspace of $L^{2}(\bbS(\Manifold),\mu_{\bbS})$.
In this paper, we deal with the cases where $\ell=1,0,-1$, and our
main concern is the case $\ell=1$.

For each $u\in\cH$, we see $u^{*}|_{\cH^{\times}}\in\CC_{1,\Manifold}^{L^{2}}$.
Hence $\cH^{*}$ is viewed as a closed subspace of $\CC_{1,\Manifold}^{L^{2}}$.

For $F:\Manifold\to\C$, define $\SYM{\tilde{F}}{Ftil}:\Proj^{-1}(\Manifold)\to\C$
by
\[
\tilde{F}(v):=F(\Proj(v)).
\]
Then we see $\tilde{F}\in\CC_{0,\Manifold}$; It follows that $\C^{\Manifold}$
(the space of functions $\Manifold\to\C$) can be identified with
$\CC_{0,\Manifold}$; We can also identify $L^{p}(\Manifold,\C)$
with $\CC_{0,\Manifold}^{L^{p}}$.

Let $F\in L^{\infty}(\Manifold,\C)$, then $F$ acts on $\CC_{\ell,\Manifold}^{L^{2}}$
as a pointwise multiplication operator $M_{F}$:
\[
\left(\SYM{M_{F}}{MF}f\right)(v):=\tilde{F}(v)f(v),\qquad v\in\Proj^{-1}(\Manifold),\ f\in\CC_{\ell,\Manifold}^{L^{2}}.
\]
Let $\SYM{E_{\cH^{*}}}{EH*}$ be the orthogonal projection from $\CC_{1,\Manifold}^{L^{2}}$
onto $\cH^{*}$. We see
\begin{equation}
E_{\cH^{*}}=\int_{\bbS(\Manifold)}\Proj(v^{*})\d\mu_{\bbS}(v).\label{eq:overcompleteS}
\end{equation}
where $\Proj(v^{*})$ is the orthogonal projection from $\CC_{1,\Manifold}^{L^{2}}$
onto $\C v^{*}$. For $f\in\CC_{1,\Manifold}^{L^{2}}$, $v\in\Proj^{-1}(\Manifold)$,
we have
\[
\left(E_{\cH^{*}}f\right)(v)=\left\langle v^{*}|E_{\cH^{*}}f\right\rangle _{\bbS}=\left\langle v^{*}|f\right\rangle _{\bbS}=\int_{\bbS(\Manifold)}\left\langle x|v\right\rangle f(x)\d\mu_{\bbS}(x).
\]
That is, the integral kernel $E_{\cH^{*}}(u,v)$ of $E_{\cH^{*}}$
is given by
\begin{equation}
E_{\cH^{*}}(v,u)=\left\langle u|v\right\rangle ,\qquad u,v\in\bbS(\Manifold).\label{eq:E-kernel}
\end{equation}

For $F\in L^{\infty}(\Manifold,\C)$, define the operator $\tilde{\cQ}(F)$
on $\CC_{1,\Manifold}^{L^{2}}$ by 

\begin{equation}
\SYM{\tilde{\cQ}(F)}{Qtil}:=\int_{\bbS(\Manifold)}\tilde{F}(v)\Proj(v^{*})\d\mu_{\bbS}(v).\label{eq:def:Qtil(F)}
\end{equation}
We see
\[
\tilde{\cQ}(F)v^{*}=\left(\cQ(F)^{*}v\right)^{*},\quad\left\langle v^{*}|\tilde{\cQ}(F)u^{*}\right\rangle _{\bbS}=\left\langle u|\cQ(F)v\right\rangle ,\qquad u,v\in\cH.
\]
{} We call the operation $\tilde{\cQ}$ the \termi{BS quantization on
$\CC_{1,\Manifold}^{L^{2}}$ }.

The following theorem says that all the information of the BS quantization
$\tilde{\cQ}$ (or $\cQ$) is essentially contained in the projection
operator $E_{\cH^{*}}$.

\begin{shaded}%
\begin{thm}
\label{thm:ProjRep}For any $F\in L^{\infty}(\Manifold,\C)$, we have
\[
\tilde{\cQ}(F)=E_{\cH^{*}}M_{F}E_{\cH^{*}}.
\]
\end{thm}

\end{shaded}
\begin{proof}
Let $u\in\cH$ and $v\in\cH^{\times}$. Then we have 
\begin{align*}
\left(E_{\cH^{*}}M_{F}E_{\cH^{*}}u^{*}\right)(v) & =\left(E_{\cH^{*}}M_{F}u^{*}\right)(v)=\left\langle v^{*}|M_{F}u^{*}\right\rangle _{\bbS}\\
 & =\int_{\bbS(\Manifold)}\d\mu_{\bbS}(x)\ol{v^{*}(x)}\tilde{F}(x)u^{*}(x)\\
 & =\int_{\bbS(\Manifold)}\d\mu_{\bbS}(x)\ol{\left\langle v|x\right\rangle }\tilde{F}(x)\left\langle u|x\right\rangle \\
 & =\int_{\bbS(\Manifold)}\d\mu_{\bbS}(x)\left\langle v^{*}|x^{*}\right\rangle _{\bbS}\left\langle x^{*}|u^{*}\right\rangle _{\bbS}\tilde{F}(x)\\
 & =\left\langle v^{*}|\tilde{\cQ}(F)u^{*}\right\rangle _{\bbS}=\left(\tilde{\cQ}(F)u^{*}\right)(v).
\end{align*}
On the other hand, if $f\in\CC_{1,\Manifold}^{L^{2}}$ is orthogonal
to $\cH^{*}$,
\[
\tilde{\cQ}(F)f=0=E_{\cH^{*}}M_{F}E_{\cH^{*}}f.
\]
\end{proof}

\subsection{Some lemmas}

Generally, for a bounded operator $A$ on $\cH$, define the operator
$\tilde{A}$ on $\CC_{1,\Manifold}^{L^{2}}$ by

\begin{align*}
\tilde{A}v^{*} & :=\left(A^{*}v\right)^{*},\quad v\in\cH,\\
\tilde{A}f & :=0\qquad\text{if }f\in\cH^{*\perp}\subset\CC_{1,\Manifold}^{L^{2}}
\end{align*}

\begin{shaded}%
\begin{lem}
\label{thm:kernel}Let $A$ be a bounded operator $A$ on $\cH$.
Define $K_{A}:\bbS(M)\times\bbS(M)\to\C$ by
\[
K_{A}(v,u)=\left\langle u|Av\right\rangle =\left\langle v^{*}|\tilde{A}u^{*}\right\rangle _{\bbS},\quad u,v\in\bbS(M).
\]
Then $K_{A}$ is the integral kernel of $\tilde{A}$, i.e.,
\[
\left(\tilde{A}f\right)(v)=\int_{\bbS(\Manifold)}\d\mu_{\bbS}(u)K_{A}(v,u)f(u),\quad\forall f\in\CC_{1,\Manifold}^{L^{2}},\ \text{a.e. }v\in\bbS(M).
\]
Especially, the kernel of $\p=\ket v\bra v\in\Manifold$ is 
\begin{equation}
K_{\p}(v',v'')=\left\langle v''|\p v'\right\rangle =\left\langle v''|v\right\rangle \left\langle v|v'\right\rangle .\label{eq:Kp()}
\end{equation}
\end{lem}

\end{shaded}
\begin{proof}
Since $\tilde{A}f=\tilde{A}E_{\cH^{*}}f$, we can choose $w\in\cH$
such that $w^{*}=E_{\cH^{*}}f$.
\begin{align*}
 & \int_{\bbS(\Manifold)}\d\mu_{\bbS}(u)\left\langle u|Av\right\rangle w^{*}(u)\\
 & \qquad=\int_{\bbS(\Manifold)}\d\mu_{\bbS}(u)\left\langle w|u\right\rangle \left\langle u|Av\right\rangle \\
 & \qquad=\left\langle w|Av\right\rangle =\left\langle A^{*}w|v\right\rangle =\left(A^{*}w\right)^{*}(v)=\left(\tilde{A}w^{*}\right)(v)=\left(\tilde{A}f\right)(v)
\end{align*}
\end{proof}

Fix $H\in L^{\infty}(\Manifold)$. For $t\in\R$, define the bounded
operator $\cQ_{t}(F)$ and $\tilde{\cQ}_{t}(F)$ on $\cH$ and $\CC_{1,\Manifold}^{L^{2}}$
respectively, by 
\[
\SYM{\cQ_{t}(F)}{Qt}:=e^{\im t\cQ(H)}\cQ(F)e^{-\im t\cQ(H)},\quad\SYM{\tilde{\cQ}_{t}(F)}{Qttil}:=e^{\im t\tilde{\cQ}(H)}\tilde{\cQ}(F)e^{-\im t\tilde{\cQ}(H)}.
\]
Note that
\begin{equation}
\tilde{\cQ}_{t}(F)=\int_{\bbS(\Manifold)}\tilde{F}(v)\Proj_{t}(v^{*})\d\mu_{\bbS}(v)=\int_{\bbS(\Manifold)}\tilde{F}(v)\Proj(v_{t}^{*})\d\mu_{\bbS}(v),\label{eq:Qtilt(F)=00003D}
\end{equation}
where
\[
\SYM{\Proj_{t}}{prt}(v^{*}):=e^{\im t\tilde{\cQ}(H)}\Proj(v^{*})e^{-\im t\tilde{\cQ}(H)},\quad v_{t}:=e^{\im t\cQ(H)}v,
\]
\[
\SYM{\p_{t}}{pt}:=e^{\im t\cQ(H)}\p e^{-\im t\cQ(H)}.
\]

\begin{shaded}%
\begin{lem}
\label{thm:Qtilt(F)sv=00003D}For $F\in L^{\infty}(\Manifold,\C)$,
\[
\left(\tilde{\cQ}_{t}(F)s\right)(v)=\int_{\bbS(\Manifold)}K_{t}(v,u)s(u)\d\mu_{\bbS}(u),\qquad s\in\CC_{1,\Manifold}^{L^{2}},
\]
where
\begin{align*}
K_{t}(v,u) & =\left\langle v^{*}|\tilde{\cQ}_{t}(F)u^{*}\right\rangle =\left\langle u|\cQ_{t}(F)v\right\rangle \\
 & =\int_{\bbS(\Manifold)}\d\mu_{\bbS}(x)\tilde{F}(x)\left\langle u|x_{t}\right\rangle \left\langle x_{t}|v\right\rangle =\int_{\Manifold}\d\mu(\p)F(\p)\left\langle u|\p_{t}v\right\rangle 
\end{align*}
\end{lem}

\end{shaded}
\begin{proof}
By (\ref{eq:Qtilt(F)=00003D}),
\begin{align*}
\left(\tilde{Q}_{t}(F)s\right)(v) & =\int_{\bbS(\Manifold)}\tilde{F}(u)\left(\Proj_{t}(u^{*})s\right)(v)\d\mu_{\bbS}(u)\\
 & =\int_{\bbS(\Manifold)}\d\mu_{\bbS}(u)\tilde{F}(u)u_{t}^{*}(v)\left\langle u_{t}^{*}|s\right\rangle _{\bbS}\\
 & =\int_{\bbS(\Manifold)}\d\mu_{\bbS}(u)\tilde{F}(u)u_{t}^{*}(v)\int_{\bbS(\Manifold)}\d\mu_{\bbS}(x)\ol{u_{t}^{*}(x)}s(x)\\
 & =\int_{\bbS(\Manifold)}\d\mu_{\bbS}(x)\left(\int_{\bbS(\Manifold)}\d\mu_{\bbS}(u)\tilde{F}(u)\left\langle x|u_{t}\right\rangle \left\langle u_{t}|v\right\rangle \right)s(x)\\
 & =\int_{\bbS(\Manifold)}\d\mu_{\bbS}(x)\left(\int_{\bbS(\Manifold)}\d\mu_{\bbS}(u)\tilde{F}(u)\left\langle v^{*}|u_{t}^{*}\right\rangle \left\langle u_{t}^{*}|x^{*}\right\rangle \right)s(x)\\
 & =\int_{\bbS(\Manifold)}\d\mu_{\bbS}(x)\left\langle v^{*}|\tilde{\cQ}_{t}(F)x^{*}\right\rangle s(x).
\end{align*}
\end{proof}

\begin{shaded}%
\begin{lem}
For $t_{1},...,t_{N}\in\R$, and $H,F_{1},...,F_{N}\in L^{\infty}(\Manifold,\C)$,
\begin{equation}
\Tr\tilde{\cQ}_{t_{1}}(F_{1})\cdots\tilde{\cQ}_{t_{N}}(F_{N})=\int_{\Manifold}\d\mu(\p_{1})\cdots\int_{\Manifold}\d\mu(\p_{N})\Tr\left(\p_{1,t_{1}}\cdots\p_{N,t_{N}}\right)\prod_{j=1}^{N}F_{j}(\p_{j}),\label{eq:TrQt1QtN}
\end{equation}
where
\[
\SYM{\p_{j,t_{j}}}{pjt}:=e^{\im t_{j}\tilde{\cQ}(H)}\p_{j}e^{-\im t_{j}\tilde{\cQ}(H)}.
\]
\end{lem}

\end{shaded}
\begin{proof}
Let $v_{N+1}:=v_{1}$. Then by Lemma \ref{thm:Qtilt(F)sv=00003D}%
,

\begin{align*}
 & \Tr\tilde{\cQ}_{t_{1}}(F_{1})\cdots\tilde{\cQ}_{t_{N}}(F_{N})\\
 & =\int_{\bbS(\Manifold)}\d\mu_{\bbS}(v_{1})\cdots\int_{\bbS(\Manifold)}\d\mu_{\bbS}(v_{N})\prod_{j=1}^{N}\int_{\Manifold}\d\mu(\p_{j})\,F_{j}(\p_{j})\left\langle v_{j}|\p_{j,t_{j}}v_{j+1}\right\rangle \\
 & =\int_{\Manifold}\d\mu(\p_{1})\cdots\int_{\Manifold}\d\mu(\p_{N})\Tr\left(\p_{1,t_{1}}\cdots\p_{N,t_{N}}\right)\prod_{j=1}^{N}F_{j}(\p_{j}).
\end{align*}
\end{proof}

\section{K\"ahler manifold}

In the following sections, we assume the
\begin{assumption}
\label{ass:Kahler}(1) $\Manifold$ is a (finite-dimensional) K\"ahler
submanifold of $\bbP\cH$, 

(2) $\Manifold$ is complete as a Riemannian manifold (or equivalently,
complete as a metric space),

(3) For any $f\in\CC_{1,\Manifold}^{L^{2}}$, $f\in\cH^{*}$ if and
only if $f$ is holomorphic.

(4) Let $\vol$ be the volume form on $\Manifold$ (as a Riemannian
manifold). Then there exists a constant $\SYM CC>0$ s.t. the measure
$\mu:=C\left|\vol\right|$ satisfies $(\Manifold,\mu)\in\Coh(\cH)$,
i.e., $\int_{\Manifold}\p\d\mu(\p)=I$.

\end{assumption}

In the following we will explain the precise meaning of these assumptions.

Note that $\bbP\cH$ has the natural topology induced by that of $\cH$,
and so $\Manifold\subset\bbP\cH$ has the topology as a subspace of
$\bbP\cH$. Assume that there exist open sets $U_{\io}\subset\C^{n}$
($\io\in\INDEX$, some index set), and holomorphic maps $\psi_{\io}:U_{\io}\to\cH^{\times}:=\cH\setminus\{0\}$
($\io\in\INDEX$) such that

(1) $\left\{ \Proj\left(\psi_{\io}(U_{\io})\right)\right\} _{\io\in\INDEX}$
is an open cover of $\Manifold$. 

(2) %
{} $\Proj\circ\psi_{\io}$ is injective for all $\io\in\INDEX$.

Then we find that $\left\{ \Proj\circ\psi_{\io}\right\} _{\io\in\INDEX}$
gives an atlas of $\Manifold$ as a complex manifold. We assume without
loss of generality that $0\in U_{\io}$ and $\left\langle \psi_{\io}(0)|\psi_{\io}(z)\right\rangle =1$
for all $z\in U_{\io}$, $\io\in\INDEX$. We write $U_{\io}':=\Proj\circ\psi_{\io}(U_{\io})\subset\Manifold$,
and locally identify $U_{\io}'$ with $U_{\io}$; We use the coordinates
$z=(z_{1},...,z_{n})$ on $U_{\io}$ also as coordinates on $U_{\io}'$.

Let $\d$ denote the exterior derivative on $\Manifold$, which is
decomposed to the holomorphic and antiholomorphic parts: $\d=\dil+\dill$.
For example, on $U_{\io}$, for $f\in C^{\infty}(U_{\io},\C)$, $\dil f$
and $\dill f$ are explicitly written as

\[
\dil f=\sum_{k}\frac{\di}{\di z_{k}}fdz_{k},\qquad\dill f=\sum_{k}\frac{\di}{\di\ol z_{k}}fd\ol z_{k}.
\]
Define $\Kpot_{\io}:U_{\io}\to\R$ by
\[
\SYM{\Kpot_{\io}}{hi}(z):=\left\Vert \psi_{\io}(z)\right\Vert ^{2},\qquad z\in U_{\io}
\]
Define the 2-form $\omega$ on $U_{\io}$by

\[
\SYM{\omega}{omega}:=-\im\,\d''\d'\log\Kpot_{\io}.
\]
It turns out that $\omega$ becomes a globally defined closed 2-form
on $\Manifold$, and hence $(\Manifold,\omega)$ is a symplectic manifold.
In fact, $\omega$ is naturally defined on whole projective space
$\bbP\cH$ as follows: Define $\Kpot:\cH^{\times}\to\R$ by $\Kpot(v):=\left\Vert v\right\Vert ^{2}$.
Then the 2-form $\omega:=-\im\,\d''\d'\log\Kpot$ on $\cH^{\times}$
is defined even when $\dim\cH=\infty$. Since $\omega$ is invariant
under the action of $\C^{\times}:=\C\setminus\{0\}$ on $\cH^{\times}$,
$\omega$ can be viewed as a 2-form on $\bbP\cH\cong\cH^{\times}/\C^{\times}$.

For tangent vector fields $X,Y$ on $\Manifold$, let
\[
g(X,Y):=\omega(X,JY),
\]
where $J$ is the complex structure on $\Manifold$. Then $g$ becomes
a Riemannian metric on $\Manifold$, and moreover we find that $(\Manifold,g,\omega)$
is a K\"ahler manifold. We call $\Kpot_{\io}$ a \termi{K\"ahler potential}
on $U_{\io}'$.

Therefore, we find that if $\Manifold$ is a (finite-dimensional)
complex submanifold of $\bbP\cH$, $\Manifold$ satisfies Assumption
\ref{ass:Kahler} (1), even when $\dim\cH=\infty$.

Assumption \ref{ass:Kahler} (2) says that every geodesic line of
$\Manifold$ can be extended for arbitrarily large values of its canonical
parameter. This is equivalent to say that $\Manifold$ is a complete
metric space with respect to the distance function $d$ induced by
the Riemannian metric $g$.

Assumption \ref{ass:Kahler} (3) says that for any $f\in\CC_{1,\Manifold}^{L^{2}}$
if $f\circ\psi_{\io}:U_{\io}\to\C$ is holomorphic for all $\io\in\INDEX$,
then $f\in\cH^{*}$. (The converse always holds.)

\subsection{Line bundle}

Let $f\in\CC_{\ell,\Manifold}$ ($\ell\in\Z$) and $v\in\Proj^{-1}(\Manifold)$.
We define the value of $f$ at $\p=\Proj(v)\in\Manifold$, denoted
$f(\p)=f(\Proj(v))$, to be

\[
\SYM{f(\p)}{f(p)}:=f|_{\ran(\p)\setminus\{0\}},\quad\text{equivalently,}\quad f(\Proj(v)):=f|_{\C^{\times}v},
\]
that is, $f(\Proj(v))$ is the function $\C^{\times}v\to\C$ defined
by
\[
f(\Proj(v))(\alpha v):=f(\alpha v)=\alpha^{\ell}f(v),\qquad\alpha\in\C^{\times}.
\]
Let
\[
\SYM{\scO_{\ell,\Manifold}}{OlM}:=\left\{ f(\p)|f\in\CC_{\ell,\Manifold},\ \p\in\Manifold\right\} ,
\]
then the natural projection $\scO_{\ell,\Manifold}\to\Manifold$,
$f(\p)\mapsto\p$ defines a complex line bundle over $\Manifold$,
where each $f\in\CC_{\ell,\Manifold}$ is a section of the line bundle.
The space $\SYM{\forms^{r}(\scO_{\ell,\Manifold})}{Ar()}$ of $\scO_{\ell,\Manifold}$-valued
$r$-forms are usually defined. For $f\in\CC_{\ell,\Manifold}$, let
${\rm supp}_{\Manifold}(f)$ denote the support of $f$ as a map $\Manifold\to\scO_{\ell,\Manifold}$,
and then for any $\alpha\in\forms^{r}(\scO_{\ell,\Manifold})$, the
support ${\rm supp}_{\Manifold}(\alpha)\subset\Manifold$ of $\alpha$
is naturally defined. 

Here, recall Lemma \ref{thm:kernel}; The integral kernel $K_{A}$
of an operator $\tilde{A}$ on $\CC_{1,\Manifold}^{L^{2}}$ was a
$\C$-valued function on $\bbS(\Manifold)\times\bbS(\Manifold)$ %
{} there. However, if $\CC_{1,\Manifold}^{L^{2}}$ is viewed as a space
of sections $s:\Manifold\to\scO_{1,\Manifold}$, the an integral kernel
$K$ of an operator on $\CC_{1,\Manifold}^{L^{2}}$ should be a map
such that $K(\p_{1},\p_{2})\in\Hom(\scO_{1,\Manifold,\p_{2}},\scO_{1,\Manifold,\p_{1}})$
for all $\p_{1},\p_{2}\in\Manifold$, where $\scO_{1,\Manifold,\p}$
is the fiber of the line bundle $\scO_{1,\Manifold}$ at $\p\in\Manifold$;
Equivalently, $K$ is a section of the external tensor product bundle
$\scO_{1,\Manifold}\boxtimes\scO_{1,\Manifold}^{*}\to\Manifold\times\Manifold$.
Note that $\scO_{1,\Manifold,\p}$ is naturally identified with the
dual space of $\ran(\p)$. %
Hence we can define $\SYM{K_{A}(\p_{1},\p_{2})}{KA(pp)}\in\Hom(\scO_{1,\Manifold,\p_{2}},\scO_{1,\Manifold,\p_{1}})$
as follows. For any $v_{2}^{*}\in\scO_{1,\Manifold,\p_{2}}=\ran(\p_{2})^{*}$
with $v_{2}\in\ran(\p_{2})$, define $K_{A}(\p_{1},\p_{2})v_{2}^{*}\in\scO_{1,\Manifold,\p_{1}}=\ran(\p_{1})^{*}$
by
\begin{equation}
\left(K_{A}(\p_{1},\p_{2})v_{2}^{*}\right)(v_{1}):=\left\langle v_{2}|Av_{1}\right\rangle =K_{A}(v_{1},v_{2}),\quad v_{1}\in\ran(\p_{1}).\label{eq:kernel-bundle}
\end{equation}
Sometimes we simply write $A(\p_{1},\p_{2})$ (resp.~$A(v_{1},v_{2})$)
for $K_{A}(\p_{1},\p_{2})$ (resp.~$K_{A}(v_{1},v_{2})$). Especially
we have
\begin{equation}
\left(E_{\cH^{*}}(\p_{1},\p_{2})v_{2}^{*}\right)(v_{1})=\left\langle v_{2}|v_{1}\right\rangle =E_{\cH^{*}}(v_{1},v_{2}),\quad v_{1}\in\ran(\p_{1}).\label{eq:kernel-E-sec}
\end{equation}
Let $f_{1},f_{2}\in\CC_{\ell,\Manifold}$ ($\ell\in\Z$). Define $\left\langle f_{1}|f_{2}\right\rangle _{{\rm pt}}:\Manifold\to\C$
by
\[
\SYM{\left\langle f_{1}|f_{2}\right\rangle _{{\rm pt}}}{<>pt}(\Proj(v)):=\ol{f_{1}\left(\frac{v}{\left\Vert v\right\Vert }\right)}f_{2}\left(\frac{v}{\left\Vert v\right\Vert }\right)=\left\Vert v\right\Vert ^{-2\ell}\ol{f_{1}\left(v\right)}f_{2}\left(v\right),\quad v\in\Proj^{-1}(\Manifold)
\]
so that if $f_{1},f_{2}\in\CC_{\ell,\Manifold}^{L^{2}}$ then
\[
\left\langle f_{1}|f_{2}\right\rangle _{\bbS}=\int_{\Manifold}\left\langle f_{1}|f_{2}\right\rangle _{{\rm pt}}C\vol.
\]
$\left\langle \cdot|\cdot\right\rangle _{{\rm pt}}$ is called the
(pointwise) \termi{Hermitian metric} of the line bundle $\scO_{\ell,\Manifold}$.

Let $\SYM{\ela}{v0}:=\psi_{\io}(0)$, then $\ela^{*}$ is a local
holomorphic section of $\scO_{1,\Manifold}$ on $U_{\io}$; Precisely,
if we identify the fiber of $\scO_{1,\Manifold}$ over $\p\in\Manifold$
with $\ran(\p)^{*}$, the value $\ela^{*}(\p)$ of the section $\ela^{*}\in\CC_{1,\Manifold}$
at $\p\in U_{\io}$ is the function $\ran(\p)\to\C$, $v\mapsto\left\langle \ela|v\right\rangle $. 

Define $h_{\io}:U_{\io}'\to\R$ by
\[
\SYM{h_{\io}(z)}{hi}:=\left\langle \ela^{*}|\ela^{*}\right\rangle _{{\rm pt}}(z),\qquad z\in U_{\io}'.
\]
Without loss of generality, assume $\left\langle \psi_{\io}(0)|\psi_{\io}(z)\right\rangle =1$,
and we find
\[
h_{\io}(z)=\left\Vert \psi_{\io}(z)\right\Vert ^{-2}\ol{\ela^{*}\left(\psi_{\io}(z)\right)}\ela^{*}\left(\psi_{\io}(z)\right)=\left\Vert \psi_{\io}(z)\right\Vert ^{-2}\ol{\left\langle \ela|\psi_{\io}(z)\right\rangle }\left\langle \ela|\psi_{\io}(z)\right\rangle =\left\Vert \psi_{\io}(z)\right\Vert ^{-2}=\frac{1}{\Kpot_{\io}(z)}.
\]
Let $\SYM{\CC_{\ell,\Manifold}^{\infty}}{GammalMinf}$ ($\ell\in\Z$)
denote the subspace of $\CC_{\ell,\Manifold}$ which consists of the
smooth functions. Any section $s\in\CC_{1,\Manifold}^{\infty}$ is
uniquely expressed locally on $U_{\io}'\subset\Manifold$ by 

\[
s(v)=f(v)\ela^{*}(v)=f(v)\left\langle \ela|v\right\rangle ,\qquad f\in\CC_{0,\Manifold}^{\infty},\ v\in\C^{\times}\psi_{\io}(U_{\io}),
\]
which is simply written as $s=f\ela^{*}$. The \termi{antiholomorphic exterior derivative}
$\dill s$ of $s=f\ela^{*}$ is defined on $U_{\io}'$ by 
\[
\dill\left(f\ela^{*}\right):=\left(\dill f\right)\otimes\ela^{*}.
\]
We find that $\dill s\in\forms^{1}(\scO_{1,\Manifold})$ is globally
well-defined on $\Manifold$. This is naturally extended to any $\scO_{1,\Manifold}$-valued
$r$-forms $\dill:\forms^{r}(\scO_{1,\Manifold})\to\forms^{r+1}(\scO_{1,\Manifold})$;
More precisely, let $\forms^{(p,q)}(\scO_{1,\Manifold})$ denote the
space of $\scO_{1,\Manifold}$-valued differential forms of type $(p,q)$,
then $\dill:\forms^{(p,q)}(\scO_{1,\Manifold})\to\forms^{(p,q+1)}(\scO_{1,\Manifold})$. 

The local \termi{Chern connection form} $\theta$ of the line bundle
$\scO_{1,\Manifold}$ on $U_{\io}\subset\C$ is defined by 
\[
\SYM{\theta}{theta}:=\d'\log h_{\io}=-\d'\log\Kpot_{\io}=\sum_{k}\left(h_{\io}^{-1}\frac{\di}{\di z_{k}}h_{\io}\right)\d z_{k}.
\]
The \termi{Chern connection} $\nabla:\forms^{0}(\scO_{1,\Manifold})\to\forms^{1}(\scO_{1,\Manifold})$
is defined as follows: for any section $s\in\CC_{1,\Manifold}^{\infty}$
which is $s=f\ela^{*}$ on $U_{\io}'\subset\Manifold$, $\nabla s$
is defined locally on $U_{\io}'$ as
\[
\SYM{\nabla}{nabla}s:=\left(\d f\right)\otimes\ela^{*}+f\nabla\ela^{*},\qquad\nabla\ela^{*}:=\theta\otimes\ela^{*}.
\]

The Chern connection $\nabla$ is Hermitian, i.e., for any vector
field $X$,%
\[
X\left\langle s_{1}|s_{2}\right\rangle _{{\rm pt}}=\left\langle \nabla_{X}s_{1}|s_{2}\right\rangle _{{\rm pt}}+\left\langle s_{1}|\nabla_{X}s_{2}\right\rangle _{{\rm pt}},\qquad s_{1},s_{2}\in\CC_{1,\Manifold}^{\infty}.
\]
Hence, for any piecewise smooth curve $c$ on $\Manifold$, the parallel
transport $\ptrans(c)$ along $c$ is unitary w.r.t.\ the Hermitian
metric $\left\langle \cdot|\cdot\right\rangle _{{\rm pt}}$. Especially,
the holonomy group at any point $\p\in\Manifold$ w.r.t. $\nabla$
is $U(1)$.%
{} Let $\ul{\psi(z)}:=\psi(z)/\left\Vert \psi(z)\right\Vert $. The
\termi{normalized frame} $\ul{\ela^{*}}\in\CC_{1,\Manifold}$ on
$U_{\io}$ is determined by 
\[
\ul{\ela^{*}}(\ul{\psi(z)})=1=h^{-1/2}(z)\ela^{*}(\ul{\psi(z)}),\qquad z\in\C,
\]
which is written $\ul{\ela^{*}}:=h_{\io}^{-1/2}\ela^{*}$ in short.
(Here, recall $\ela^{*}(\psi_{\io}(z))=\left\langle \psi_{\io}(0)|\psi_{\io}(z)\right\rangle =1$
and $h_{\io}(z)^{-1}=\left\Vert \psi_{\io}(z)\right\Vert ^{2}=\Kpot_{\io}(z).$)
Precisely, the value $\ul{\ela^{*}}(\p)$ of the section $\ul{\ela^{*}}$
at $\p\in U_{\io}$ is the function
\[
\ran(\p)\to\C,\qquad\zeta\ul{\psi(z)}\mapsto h_{\io}^{-1/2}(z)\bigl\langle\ela^{*}|\zeta\ul{\psi(z)}\bigr\rangle=\zeta,\qquad z\in\C^{d},\ \psi(z)\in\ran(\p),\ \zeta\in\C.
\]
For $s=f\ul{\ela}^{*}$, $f\in C^{\infty}(U_{\io},\C)$, we have
\[
\nabla s=\left(\d f\right)\otimes\ul{\ela}^{*}+f\im\theta_{{\rm nor}}\otimes\ul{\ela}^{*},
\]
where
\begin{equation}
\SYM{\theta_{{\rm nor}}}{thetanor}:=\frac{1}{2}\sum_{k}\left[\left(\frac{\di}{\di x_{k}}\log h_{\io}\right)\d y_{k}-\left(\frac{\di}{\di y_{k}}\log h_{\io}\right)\d x_{k}\right],\quad z_{k}=x_{k}+\im y_{k},\label{eq:def:thetanor}
\end{equation}
which is a $\R$-valued 1-form; Since $\im\R=\fraku(1)$ (the Lie
algebra of $U(1)$), this says that the Chern connection $\nabla$
determines a connection on the associated $U(1)$ principal bundle,
explicitly written by $\im\theta_{{\rm nor}}$.

For $s=f\ela^{*}$, let%
\[
\SYM{\nabla'}{nabla'}s:=\left(\d^{\prime}f\right)\otimes\ela^{*}+f\left(\theta\otimes\ela^{*}\right),
\]
then we have 
\[
\nabla=\nabla'+\dill.
\]
We find that $\nabla:\forms^{0}(\scO_{1,\Manifold})\to\forms^{1}(\scO_{1,\Manifold})$
and $\nabla':\forms^{(0,0)}(\scO_{1,\Manifold})\to\forms^{(1,0)}(\scO_{1,\Manifold})$
are globally well-defined on $\Manifold$. 

For general differential forms $\alpha\in\forms^{r}(\scO_{1,\Manifold})$,
$\nabla\alpha\in\forms^{r+1}(\scO_{1,\Manifold})$ is defined by 
\[
\SYM{\nabla}{nabla}\alpha:=\sum_{j}(\d\alpha_{j}\otimes s_{j}+(-1)^{r}\alpha_{j}\wedge\nabla s_{j})
\]
where $\alpha={\displaystyle \sum_{j}\alpha_{j}\otimes s_{j}}$ ($\alpha_{j}\in\forms^{r}(\Manifold,\C)$,
$s_{j}\in\CC_{1,\Manifold}^{\infty}$).

It is shown that there exists $\Theta\in\forms^{2}(\Manifold,\C)$
such that
\begin{equation}
\nabla^{2}\alpha=\Theta\wedge\alpha,\qquad\forall\alpha\in\forms^{k}(\scO_{1,\Manifold}),\ k=0,...,n.\label{eq:Ra=00003DRwedgea}
\end{equation}
$\Theta$ is called the \termi{curvature form} of the Chern connection
$\nabla$. Generally, the curvature form $\Theta$ is locally expressed
by the Hermitian metric $h$ by 
\[
\Theta=\d''\d'\log h_{\io},
\]
and so we have
\begin{equation}
\omega=\im\Theta\label{eq:omega=00003DiTheta}
\end{equation}
in our case.%

\subsection{The Bochner-Kodaira-Nakano identity}

Let $\CC_{1,\Manifold,{\rm c}}^{\infty}$ denote the space of compactly
supported smooth sections of $\scO_{1,\Manifold}$, and $\forms_{{\rm c}}^{r}(\scO_{1,\Manifold})$
the space of compactly supported $\scO_{1,\Manifold}$-valued $r$-forms
on $\Manifold$. The inner product $\left\langle \cdot|\cdot\right\rangle _{\bbS}$
is naturally extended to $\forms_{{\rm c}}^{r}(\scO_{1,\Manifold})$,
and the formal adjoint operators are usually defined on these spaces:
$\nabla^{*}$, $\nabla^{\prime*}$ and $\dill^{*}$are the formal
adjoints of $\nabla$, $\nabla'$ and $\dill$, respectively. We use
the notation $\SYM{\Delta_{X}}{DeltaX}:=X^{*}X+XX^{*}$ for any operator
$X$ on $\forms_{c}^{r}(\scO_{1,\Manifold})$, if the formal adjoint
$X^{*}$ of $X$ exists. For example,
\[
\Delta_{\dill}:=\dill^{*}\dill+\dill\dill^{*},\qquad\Denabla:=\nabla^{*}\nabla+\nabla\nabla^{*}.
\]
The \termi{Lefschetz map} $L:\forms^{(p,q)}(\scO_{1,\Manifold})\to\forms^{(p+1,q+1)}(\scO_{1,\Manifold})$
is defined by
\[
\SYM{L(\alpha)}{L(alp)}:=\omega\wedge\alpha.
\]
Since $\omega=\im\Theta$ holds in our case, we have $L(\alpha)=\im\Theta\wedge\alpha=\im\nabla^{2}\alpha$,
i.e.,
\begin{equation}
L=\im\nabla^{2}.\label{eq:L=00003Dinabla2}
\end{equation}
Let $\SYM{P_{r}}{Pr}$ denote the natural projection onto $\forms^{r}(\scO_{1,\Manifold})$:
If $\alpha=\sum_{k=0}^{\dim\Manifold}\alpha_{k}$, $\alpha_{k}\in\forms^{k}(\scO_{1,\Manifold})$,
then $P_{r}(\alpha):=\alpha_{r}$.

\begin{shaded}%
\begin{lem}
\label{thm:=00005BL*L=00005D}(See e.g. \cite[Ch.5]{Bal06}; \cite[Eq.(3.2.37)]{Kob87})
{} $[L^{*},L]=\sum_{r=0}^{n}(n-r)P_{r}$, where $n=\dim_{\R}\Manifold$.
\end{lem}

\end{shaded}

\begin{shaded}%
\begin{thm}
{\rm (\termi{Bochner--Kodaira--Nakano identity})} (See e.g. \cite[Ch.5]{Bal06})%
{} %
\begin{equation}
2\Delta_{\d''}=\Denabla+[\im\nabla^{2},L^{*}]\label{eq:2Ded''=00003DDed}
\end{equation}
\end{thm}

\end{shaded}

By (\ref{eq:L=00003Dinabla2}): $L=\im\nabla^{2}$, Lemma \ref{thm:=00005BL*L=00005D}
{} and the Bochner--Kodaira--Nakano identity (\ref{eq:2Ded''=00003DDed}),
we have

\begin{shaded}%
\begin{cor}
\label{thm:tautDelta} On the line bundle $\scO_{1,\Manifold}$ over
$\Manifold$, we have\textup{
\[
2\Delta_{\d''}=\Denabla+[L,L^{*}]=\Denabla-\sum_{r=0}^{n}(n-r)P_{r},\qquad n=\dim_{\R}\Manifold.
\]
}%
\textup{Especially,
\[
2\Delta_{\d''}s=\left(\Denabla-n\right)s=\left(\nabla^{*}\nabla-n\right)s,\qquad\forall s\in\forms^{0}(\scO_{1,\Manifold})=\CC_{1,\Manifold}^{\infty}.
\]
}
\end{cor}

\end{shaded}

Generally, the operators $\Denabla$ and $\Delta_{\d''}$ are shown
to be essentially self-adjoint, and we write the self-adjoint extensions
these operators again by $\Denabla$ and $\Delta_{\d''}$, respectively.

\begin{shaded}%
\begin{cor}
\[
E_{\cH^{*}}=\slim_{\no\to\infty}\exp\left[-\no\left(\nabla^{*}\nabla-n\right)\right],
\]
where $\slim$ denotes the limit in the strong topology on $B(\CC_{1,\Manifold}^{L^{2}})$,
the space of bounded operators on $\CC_{1,\Manifold}^{L^{2}}$. 
\end{cor}

\end{shaded}

We say that $\Delta_{\d''}$ has a \termi{spectral gap} if $\inf\left({\rm spec}(\Delta_{\d''})\setminus\{0\}\right)>0$,
where ${\rm spec}(\Delta_{\d''})$ is the spectrum of $\Delta_{\d''}$;
In other words, $\Delta_{\d''}\ge\alpha(I-E_{\cH^{*}})$ for some
$\alpha>0$.

\begin{shaded}%
\begin{cor}
If $\Delta_{\d''}$ has a spectral gap, then 
\[
E_{\cH^{*}}=\lim_{\no\to\infty}\exp\left[-\no\left(\nabla^{*}\nabla-n\right)\right],
\]
where $\lim$ denotes the limit in operator norm on $B(\CC_{1,\Manifold}^{L^{2}})$.
\end{cor}

\end{shaded}

\section{Asymptotic representation}

Let $\cL$ be an arbitrary complex Hilbert space, and $\cK$ be a
closed subspace of $\cL$. Let $H$ be a bounded self-adjoint operator
on $\cL$, and $A$ a possibly unbounded positive semidefinite operator
on $\cL$ such that $\ker A=\cK\subset\cL$. Assume that $A$ has
a spectral gap, i.e., there exists $\alpha>0$ such that the spectrum
of $A$ satisfies $\spec(A)\cap(0,\alpha)=\emptyset$, equivalently
$A^{2}\ge\alpha A$, or $A\ge\alpha E_{\cK}^{\perp}$ where $E_{\cK}^{\perp}:=1-E_{\cK}$.

This section we show the following theorem:

\begin{shaded}%
\begin{thm}
\label{thm:sibori} For all $t\ge0$,
\begin{align*}
\lim_{\no\to\infty}e^{-t\left(\no A+\im H\right)}v & =e^{-\im tE_{\cK}HE_{\cK}}v,\qquad\forall v\in\cK=\ker A,\\
\lim_{\no\to\infty}e^{-t\left(\no A+\im H\right)}v & =0,\qquad\forall v\in\cK^{\perp}.
\end{align*}
That is,
\[
\slim_{\no\to\infty}e^{-t\left(\no A+\im H\right)}=e^{-\im tE_{\cK}HE_{\cK}}E_{\cK}=E_{\cK}e^{-\im tE_{\cK}HE_{\cK}}E_{\cK}.
\]
\end{thm}

\end{shaded}

Note that we can give a precise meaning to the operator $e^{-t\left(\no A+\im H\right)}$
as follows. We find that $T_{\no}:=\no A+\im H$ is a closed operator
satisfying $\Re\langle v|T_{\no}v\rangle\ge0$ for all $v\in\dom(T_{\no})=\dom(A)$.
Hence $T_{\no}$ generates the strongly continuous contraction semigroup
$\{e^{-tT_{\no}}|t\ge0\}$ by the Hille--Yosida Theorem \cite{RS75}. 

Recall the notations and assumptions in the previous section. Set
\[
\cL:=\CC_{1,\Manifold}^{L^{2}},\quad\cK:=\cH^{*}\subset\CC_{1,\Manifold}^{L^{2}},\quad A:=2\Delta_{\d''}\upha_{\CC_{1,\Manifold}^{L^{2}}}=\nabla^{*}\nabla-n,\quad n:=\dim_{\R}\Manifold,
\]
in Theorem \ref{thm:sibori}, then we have the following:

\begin{shaded}%
\begin{cor}
\label{thm:siboriQ} Assume that $\Delta_{\d''}$ has a spectral gap.
Then for any ``classical Hamiltonian'' $H\in L^{\infty}(\Manifold,\R)$,
the ``quantum time evolution'' $e^{-\im t\tilde{\cQ}(H)}$ is expressed
as
\[
e^{-\im t\tilde{\cQ}(H)}E_{\cH^{*}}=E_{\cH^{*}}e^{-\im t\tilde{\cQ}(H)}E_{\cH^{*}}=\slim_{\no\to\infty}\exp\left[-t\left(\no(\nabla^{*}\nabla-n)+\im H\right)\right],\quad t\ge0,
\]
where $H$ in the rhs is viewed as a multiplication operator $M_{H}$
on $\CC_{1,\Manifold}^{L^{2}}$.
\end{cor}

\end{shaded}

A similar statement and its proof are found in \cite{Yam11}, but
that proof was somewhat erroneous, and confusing in notations. Thus
we will present another proof here. To complete the proof, we need
some lemmas.

\begin{lyxgreyedout}
\hidec{

\begin{shaded}%
\begin{lem}
Let $H$ be a bounded self-adjoint operator on $\cL$, and $A$ a
possibly unbounded positive operator on $\cL$ such that $\ker A=\cK\subset\cL$.
Let $E_{\cK}^{\perp}:=1-E_{\cK}$, and assume that there exists $\alpha>0$
such that $A\ge\alpha E_{\cK}^{\perp}$. %
{} Then if $v\in\dom(A)$ satisfies 
\[
\frac{d}{dt}\Big|_{t=0}\left\Vert E_{\cK}^{\perp}e^{-t\left(\nu A+\im H\right)}v\right\Vert ^{2}\ge0,
\]
we have 
\[
\left\Vert E_{\cK}^{\perp}v\right\Vert \le\frac{\left\Vert H\right\Vert \left\Vert v\right\Vert }{\alpha^{2}\nu}.\ 
\]
\end{lem}

\end{shaded}
\begin{proof}
\end{proof}
\rule[0.5ex]{1\columnwidth}{1pt}

\begin{shaded}%
\begin{lem}
\label{thm:Ev<Hval2-1}Let $H$ be a bounded self-adjoint operator
on $\cL$, and $A$ a possibly unbounded positive semidefinite operator
on $\cL$ such that $\ker A=\cK\subset\cL$. Let $E_{\cK}^{\perp}:=1-E_{\cK}$,
and assume that there exists $\alpha>0$ such that $A\ge\alpha E_{\cK}^{\perp}$.
{} Then if $v\in\dom(A)$ satisfies
\[
\frac{d}{dt}\Big|_{t=0}\left\Vert E_{\cK}^{\perp}e^{-t\left(A+\im H\right)}v\right\Vert ^{2}\ge0,
\]
we have $\left\Vert E_{\cK}^{\perp}v\right\Vert \le\left\Vert H\right\Vert \left\Vert v\right\Vert /\alpha^{2}.$
\end{lem}

\end{shaded}
\begin{proof}
\begin{lyxgreyedout}
\begin{proof}
By the equation $\frac{d}{dt}|_{t=0}\left\Vert \varphi(t)\right\Vert ^{2}=2\Re\left\langle \varphi(0)|\dot{\varphi}(0)\right\rangle $,
we have
\[
\frac{d}{dt}|_{t=0}\left\Vert E_{\cK}^{\perp}e^{-t\left(A+\im H\right)}v\right\Vert ^{2}=2\Re\left\langle E_{\cK}^{\perp}e^{-t\left(A+\im H\right)}v|\frac{d}{dt}E_{\cK}^{\perp}e^{-t\left(A+\im H\right)}v\right\rangle |_{t=0}\ 
\]
\[
=2\Re\left\langle E_{\cK}^{\perp}v|E_{\cK}^{\perp}\left(-\left(A+\im H\right)\right)v\right\rangle \ 
\]
\[
=-2\Re\left\langle E_{\cK}^{\perp}v|\left(A+\im E_{\cK}^{\perp}H\right)v\right\rangle \ 
\]
\[
=-2\Re\left\langle v|\left(A+\im E_{\cK}^{\perp}H\right)v\right\rangle \ 
\]
\[
=-2\Re\left[\left\langle v|Av\right\rangle +\left\langle v|\left(\im E_{\cK}^{\perp}H\right)v\right\rangle \right]\ 
\]

\[
=-2\Re\left[\left\langle v|Av\right\rangle +\im\left\langle E_{\cK}^{\perp}v|Hv\right\rangle \right]\ 
\]
\[
=-2\left[\left\langle v|Av\right\rangle +\Re\im\left\langle E_{\cK}^{\perp}v|Hv\right\rangle \right]\ 
\]
Hence
\[
\frac{d}{dt}|_{t=0}\left\Vert E_{\cK}^{\perp}e^{-t\left(A+\im H\right)}v\right\Vert ^{2}\ge0\iff-2\left[\left\langle v|Av\right\rangle +\Re\im\left\langle E_{\cK}^{\perp}v|Hv\right\rangle \right]\ge0
\]
\[
\iff\left\langle v|Av\right\rangle \le-\Re\im\left\langle E_{\cK}^{\perp}v|Hv\right\rangle 
\]
Since $A\ge\alpha E_{\cK}^{\perp}$, $\left\langle v|Av\right\rangle \ge\left\langle v|\alpha E_{\cK}^{\perp}v\right\rangle =\alpha^{2}\left\Vert E_{\cK}^{\perp}v\right\Vert ^{2}$,
\[
\then\alpha^{2}\left\Vert E_{\cK}^{\perp}v\right\Vert ^{2}\le-\Re\im\left\langle E_{\cK}^{\perp}v|Hv\right\rangle \ 
\]
$-\Re\im\left\langle E_{\cK}^{\perp}v|Hv\right\rangle \le\left|\left\langle E_{\cK}^{\perp}v|Hv\right\rangle \right|\le\left\Vert E_{\cK}^{\perp}v\right\Vert \left\Vert Hv\right\Vert \le\left\Vert E_{\cK}^{\perp}v\right\Vert \left\Vert H\right\Vert _{\infty}\left\Vert v\right\Vert $
\[
\then\alpha^{2}\left\Vert E_{\cK}^{\perp}v\right\Vert ^{2}\le\left\Vert E_{\cK}^{\perp}v\right\Vert \left\Vert H\right\Vert _{\infty}\left\Vert v\right\Vert \ 
\]
\[
\then\left\Vert E_{\cK}^{\perp}v\right\Vert \le\alpha^{-2}\left\Vert H\right\Vert _{\infty}\left\Vert v\right\Vert \ 
\]
\end{proof}
\end{lyxgreyedout}
\end{proof}
\rule[0.5ex]{1\columnwidth}{1pt}

}%
\end{lyxgreyedout}

\begin{shaded}%
\begin{lem}
\label{thm:vAv<Hv/al}%
{} Let $v\in\cL$ and $v_{t}:=e^{-t\left(A+\im H\right)}v$, $t\ge0$.
If $v_{t}\in\dom A$ and $\frac{d}{dt}\left\langle v_{t}|Av_{t}\right\rangle \ge0$,
then we have 
\[
\left\Vert Av_{t}\right\Vert \le\left\Vert H\right\Vert \left\Vert v_{t}\right\Vert ,\quad\text{and}\quad\left\langle v_{t}|Av_{t}\right\rangle \le\alpha^{-1}\left\Vert H\right\Vert ^{2}\left\Vert v_{t}\right\Vert ^{2}.
\]
\end{lem}

\end{shaded}
\begin{proof}
We see $\frac{d}{dt}\left\langle v_{t}|Av_{t}\right\rangle =-2\left(\left\langle Av_{t}|Av_{t}\right\rangle +\Re\left\langle \im Hv_{t}|Av_{t}\right\rangle \right)$%
{} and $\left|\left\langle \im Hv_{t}|Av_{t}\right\rangle \right|\le\left\Vert H\right\Vert \left\Vert v_{t}\right\Vert \left\Vert Av_{t}\right\Vert $.
Hence $\frac{d}{dt}\left\langle v_{t}|Av_{t}\right\rangle \ge0$ implies
$\left\Vert Av_{t}\right\Vert ^{2}\le\left\Vert H\right\Vert \left\Vert v_{t}\right\Vert \left\Vert Av_{t}\right\Vert $,
i.e. $\left\Vert Av_{t}\right\Vert \le\left\Vert H\right\Vert \left\Vert v_{t}\right\Vert $.
Since $A^{2}\ge\alpha A$, we have $\|\left(\alpha A\right)^{1/2}v_{t}\|\le\left\Vert H\right\Vert \left\Vert v_{t}\right\Vert $.
This is equivalent to $\left\langle v_{t}|Av_{t}\right\rangle \le\alpha^{-1}\left\Vert H\right\Vert ^{2}\left\Vert v_{t}\right\Vert ^{2}$.%
\end{proof}
\begin{shaded}%
\begin{lem}
\label{thm:Av<Hv}%
{} Let $v\in\cK=\ker A$, and $v_{t}:=e^{-t\left(A+\im H\right)}v$,
$t\ge0$. Then for all $t\ge0$, we have $v_{t}\in\dom A$ and
\begin{equation}
\left\Vert Av_{t}\right\Vert \le\left\Vert H\right\Vert \left\Vert v\right\Vert ,\quad\left\langle v_{t}|Av_{t}\right\rangle \le\frac{\left\Vert H\right\Vert ^{2}\left\Vert v\right\Vert ^{2}}{\alpha},\label{eq:Av<Hv}
\end{equation}
\begin{equation}
\left\Vert E_{\cK}^{\perp}v_{t}\right\Vert \le\frac{\left\Vert H\right\Vert \left\Vert v\right\Vert }{\alpha}.\label{eq:Ev<Hv}
\end{equation}
\end{lem}

\end{shaded}
\begin{proof}
The first and second inequalities are the direct consequences of Lemma
\ref{thm:vAv<Hv/al}. Recall $\alpha^{-1}A\ge E_{\cK}^{\perp}$, then
the third follows from
\[
\left\Vert E_{\cK}^{\perp}v_{t}\right\Vert ^{2}=\left\langle v_{t}|E_{\cK}^{\perp}v_{t}\right\rangle \le\alpha^{-1}\left\langle v_{t}|Av_{t}\right\rangle \le\alpha^{-2}\left\Vert H\right\Vert ^{2}\left\Vert v\right\Vert ^{2}.
\]
\end{proof}
\begin{lyxgreyedout}
\hidec{

\begin{shaded}%
$v_{t}:=e^{-t\left(A+\im H\right)}v$,
\begin{equation}
\left\Vert \left(\frac{d}{dt}+\im E_{\cK}HE_{\cK}\right)E_{\cK}v_{t}\right\Vert \le\frac{\left\Vert H\right\Vert ^{2}\left\Vert v\right\Vert }{\alpha}\label{eq:E(dv+iEHEv)<-1}
\end{equation}
\end{shaded}
\begin{proof}
\[
\left(\frac{d}{dt}+\im E_{\cK}HE_{\cK}\right)E_{\cK}v_{t}=\frac{d}{dt}E_{\cK}v_{t}+\im E_{\cK}HE_{\cK}v_{t}
\]
\[
=E_{\cK}\left(-\left(A+\im H\right)\right)v_{t}+\im E_{\cK}HE_{\cK}v_{t}
\]

\[
=-\im E_{\cK}Hv_{t}+\im E_{\cK}HE_{\cK}v_{t}=-\im E_{\cK}H\left(1-E_{\cK}\right)v_{t}=-\im E_{\cK}HE_{\cK}^{\perp}v_{t}
\]
\[
\left\Vert \left(\frac{d}{dt}E_{\cK}+\im E_{\cK}HE_{\cK}\right)v_{t}\right\Vert =\left\Vert E_{\cK}HE_{\cK}^{\perp}v_{t}\right\Vert \le\left\Vert H\right\Vert \left\Vert E_{\cK}^{\perp}v_{t}\right\Vert \ \ 
\]
\[
\le_{\footnotesize(\ref{eq:Ev<Hv})}\left\Vert H\right\Vert \frac{\left\Vert H\right\Vert \left\Vert v\right\Vert }{\alpha}=\frac{\left\Vert H\right\Vert ^{2}\left\Vert v\right\Vert }{\alpha}\ 
\]
\end{proof}
\rule[0.5ex]{1\columnwidth}{1pt}

}%
\end{lyxgreyedout}

\noindent{}\textit{Proof of Theorem \ref{thm:sibori}}: Substitute
$\no A$ for $A$, and so $\no\alpha$ for $\alpha$ in the above
lemmas, and let $v\in\cK$ and
\[
v_{t}:=v_{t}^{(\no)}:=e^{-t\left(\no A+\im H\right)}v
\]
We find
\[
\left(\frac{d}{dt}+\im E_{\cK}HE_{\cK}\right)E_{\cK}v_{t}^{(\no)}=-\im E_{\cK}HE_{\cK}^{\perp}v_{t}^{(\no)},
\]
and hence
\[
\left\Vert \left(\frac{d}{dt}+\im E_{\cK}HE_{\cK}\right)E_{\cK}v_{t}^{(\no)}\right\Vert =\left\Vert E_{\cK}HE_{\cK}^{\perp}v_{t}^{(\no)}\right\Vert \le\left\Vert H\right\Vert \left\Vert E_{\cK}^{\perp}v_{t}^{(\no)}\right\Vert \ \ 
\]
\begin{equation}
\le_{\footnotesize(\ref{eq:Ev<Hv})}\left\Vert H\right\Vert \frac{\left\Vert H\right\Vert \left\Vert v\right\Vert }{\no\alpha}=\frac{\left\Vert H\right\Vert ^{2}\left\Vert v\right\Vert }{\no\alpha}\ \label{eq:(ddt+iEHE)Ev<}
\end{equation}
Thus we have
\[
\lim_{n\to\infty}\left\Vert \left(\frac{d}{dt}+\im E_{\cK}HE_{\cK}\right)E_{\cK}v_{t}^{(\no)}\right\Vert =0.
\]
This implies $u_{t}:=\lim_{\no\to\infty}E_{\cK}v_{t}^{(\no)}$ exists,
and satisfies
\[
\left(\frac{d}{dt}+\im E_{\cK}HE_{\cK}\right)u_{t}=0,\quad\text{i.e.}\quad u_{t}=e^{-\im tE_{\cK}HE_{\cK}}v.
\]
On the other hand, again by (\ref{eq:Ev<Hv})%
\[
\lim_{n\to\infty}\left\Vert E_{\cK}^{\perp}v_{t}^{(\no)}\right\Vert \le\lim_{\no\to\infty}\frac{\left\Vert H\right\Vert \left\Vert v\right\Vert }{\no\alpha}=0.
\]
Hence we have
\[
\lim_{\no\to\infty}v_{t}^{(\no)}=\lim_{\no\to\infty}E_{\cK}v_{t}^{(\no)}=u_{t}.
\]
\qed

\section{Path integral representation}

In this section, first we state a Feynman--Kac formula on a Riemannian
manifold. Our main source is G\"uneysu \cite{Gun10}.

Let $(M,g)$ be a complete Riemannian manifold. The Laplace-Beltrami
operator on $M$ is denoted with
\[
\SYM{\Delta}{Delta}=-\mathrm{d}^{*}\mathrm{d}:C^{\infty}(M)\rightarrow C^{\infty}(M).
\]
The following fact is known (see e.g.\ \cite{Gun10}, Theorem 2.24):

\begin{shaded}%
\begin{thm}
{} There exists a unique minimal positive fundamental solution
\[
p:(0,\infty)\times M\times M\rightarrow(0,\infty),\quad(t,x,y)\mapsto p_{t}(x,y)
\]
of the heat equation
\[
\frac{\partial}{\partial t}h(t,x)=\frac{1}{2}\Delta_{x}h(t,x).
\]
\qed
\end{thm}

\end{shaded}

The map $p$ is called the \termi{minimal heat kernel} of $M=(M,g)$,
\begin{defn}
{} Let $\SYM XX$ be a continuous semi-martingale with values in $M$
in the time interval ${\bf T}=[t_{0},t_{1}]$ or $[t_{0},\infty)$.
Fix a smooth principal bundle $\pi:\SYM PP\rightarrow M$ with structure
group $\SYM GG$ and the associated Lie algebra $\SYM{\fg}g$, and
a connection 1-form $\SYM{\alpha_{0}}{alpha0}\in\forms^{1}(P,\fg)$.
A continuous semi-martingale $U$ on $P$ defined in the time interval
${\bf T}$ is called a \termi{horizontal lift} of $X$ to $P$ (with
respect to the connection $\alpha_{0}$), if $\pi(U)=X$ and
\[
\int\alpha_{0}(\underline{\mathrm{d}}U)=0,
\]
where $\ul{\d}$ denotes the Stratonovich integral.
\end{defn}

\begin{shaded}%
\begin{prop}
(\cite{Gun10}, Theorem 2.15) There is a unique horizontal lift $U$
of $X$ to $P$ with $U_{0}=u_{0}$ $\mathbb{P}$-a.s.\qed
\end{prop}

\end{shaded}

Let $E\to M$ be a Hermitian vector bundle with a fixed Hermitian
connection $\nabla$. Let $\SYM{\pi}{pi}:\mathrm{P}(E)\rightarrow M$
be the $\mathrm{U}(d)$-principal bundle corresponding to $(E,(\bullet,\bullet)_{x})$,
that is,
\[
\SYM{\mathrm{P}(E)}{P(E)}=\bigcup_{x\in M}\{u|u:\C^{d}\stackrel{\simeq}{\longrightarrow}E_{x}\text{ is an isometry}\}.
\]
The Hermitian connection $\nabla$ on the vector bundle $E\to M$
is induced by a unique connection 1-form $\alpha_{0}\in\forms^{1}(P(E),\fg)$
on $P$, $\fg:=\fraku(d)$.

Let $\SYM{E\boxtimes E^{*}}{E\boxtimes E^{*}}\rightarrow M\times M$
be the external tensor product bundle corresponding to $E,$ that
is,
\[
E\boxtimes E^{*}|_{(x,y)}=E_{x}\otimes E_{y}^{*}=\Hom(E_{y},E_{x}).
\]

\begin{shaded}%
\begin{prop}
(\cite{Gun10}, Proposition and definition 2.17) Let $X$ be a continuous
semi-martingale with values in $M$ in the time interval %
{} ${\bf T}=[t_{0},t_{1}]$ or $[t_{0},\infty)$. Let $U$ be a horizontal
lift of $X$ to $\mathrm{P}(E)$ w.r.t.\ the connection 1-form $\alpha_{0}$.
Then the continuous adapted process given by
\[
\SYM{\ptrans^{X}}{//X}:=UU_{0}^{-1}:{\bf T}\times\Omega\rightarrow E\boxtimes E^{*}
\]
does not depend on the particular choice of the lift $U$, and $\ptrans^{X}$
is called the \termi{stochastic parallel transport} in $E$ along
$X.$\qed
\end{prop}

\end{shaded}

Let $V\in L_{\loc}^{2}(M,\R)$ and assume that $V$ is bounded from
below, and is in the local Kato class. Define a self-adjoint operator

\[
\SYM{H(V)}{H(V)}:=\frac{1}{2}\nabla^{*}\nabla+V
\]
on $\Gamma_{L^{2}}(M,E)$, the Hilbert space of $L^{2}$ sections
of $E$; Precisely, $\nabla^{*}\nabla/2+V$ is shown to be essentially
self-adjoint on the domain of smooth sections with compact support,
and $H(V)$ denotes the self-adjoint extension of it.

\begin{shaded}%
\begin{prop}
\label{thm:bridge}If $M$ is geodesically complete with Ricci curvature
bounded from below and a positive injectivity radius, then there exists
the Brownian bridge measures $\mathbb{P}_{t}^{x,y}$ in a way that
the expectation values $\Ex_{t}^{x,y}[\bullet]$ are a rigorous version
of the conditional expectation values $\Ex[\bullet|B_{t}(x)=y]$.\qed
\end{prop}

\end{shaded}

\begin{shaded}%
\begin{thm}
\label{thm:Gun1.2}(\cite{Gun10}, Theorem 1.2) Let $M$ satisfy the
conditions of Prop. \ref{thm:bridge}, and $B$ a Brownian motion
on $\Manifold$ in the time interval $[0,\infty)$. For any $t>0$,
define the section $K_{t}$ of the bundle $E\boxtimes E^{*}\to M\times M$,
i.e., $K_{t}(x,y)\in\Hom(E_{y},E_{x})$ for all $x,y\in M$, by%
\begin{equation}
K_{t}(x,y):=p_{t}(x,y)\Ex_{t}^{x,y}[\scV_{t}^{B}\ptrans_{t}^{B,-1}],\quad\SYM{\scV_{t}^{B}}{Valtx}:=e^{-\int_{0}^{t}V(B_{s})\mathrm{d}s},\quad x,y\in M.\label{eq:Kt(x,y)=00003DptEt}
\end{equation}
Then $K_{t}$ defines an bounded integral kernel for the operator
$e^{-tH(V)}.$ We write $e^{-tH(V)}(x,y):=K_{t}(x,y)$.\qed
\end{thm}

\end{shaded}

Here, $V$ is assumed to be real-valued. We want to extend this formula
to the cases where $V$ is complex-valued. However, in that case $H(V)$
is not self-adjoint, and the analysis appears to become far more difficult.
The extension is easy only when $V$ is bounded, and its proof is
similar to the case where $V$ is real-valued. We will consider this
case in the following.

For each $x,y\in M$ and $0\le t_{0}<t_{1}$, let $\SYM{C_{x,y}([t_{0},t_{1}],M)}{Cxy}$
denote the space of continuous functions $\path:[t_{0},t_{1}]\to M$
such that $\path(t_{0})=x$ and $\path(t_{1})=y$. Then we can choose
$C_{x,y}([0,t],M)$ (or more generally $C_{x,y}([t_{0},t_{0}+t],M)$
as a natural sample space $\Omega$ of the probability measure $\mathbb{P}_{t}^{x,y}$.
It follows from Theorem \ref{thm:Gun1.2} that there exists a finite
measure $\mu_{[0,t];x,y}:=p_{t}(x,y)\mathbb{P}_{t}^{x,y}$ on $C_{x,y}([0,t],M)$
such that 
\begin{equation}
e^{-tH(V)}(x,y)=\int_{C_{x,y}([0,t],M)}e^{-\int_{0}^{t}V(\path_{t})\mathrm{d}s}\ptrans_{t}^{\path,-1}\d\mu_{[0,t];x,y}(\path).\label{eq:e^itH(V)(x,y)=00003Ddmu}
\end{equation}
In the following we use this form of path integral formula instead
of the probabilistic form (\ref{eq:Kt(x,y)=00003DptEt}), mainly because
the appearance of (\ref{eq:e^itH(V)(x,y)=00003Ddmu}) is more similar
to physicists' naive path integral formulas than that of (\ref{eq:Kt(x,y)=00003DptEt}).
{} %

Recall the definitions and notations in previous sections, together
with Assumption \ref{ass:Kahler}; Set $M:=\Manifold$, and $E$ to
be the Hermitian line bundle $\scO_{1,\Manifold}$ over $\Manifold$.
Let $\nabla$ be the Chern connection on $\scO_{1,\Manifold}$. 

For each $\no>0$, consider the measure $\mu_{0,\no t;x,y}$ on $C_{x,y}([0,\no t],\Manifold)$.
Let $\SYM{\path_{t}^{(\no)}}{Xt(no)}:=\path_{\no t}$, and define
the measure $\SYM{\mu_{0,t;x,y}^{(\no)}}{mu(no)}$ on $C_{x,y}([0,t],\Manifold)$
by the time rescaling $t\mapsto\no t$: 
\[
\d\mu_{0,t;x,y}^{(\no)}(\path^{(\no)}):=\d\mu_{0,\no t;x,y}(\path).
\]

{} %

\begin{shaded}%
\begin{thm}
\label{thm:Ut(pp)=00003D}Let $\Manifold$ satisfy Assumption \ref{ass:Kahler}
and the conditions of Proposition \ref{thm:bridge} as a Riemannian
manifold. Furthermore assume that the operator $\Delta_{\d''}$ on
the line bundle $\scO_{1,\Manifold}$ has a spectral gap. Let $H\in L^{\infty}(\Manifold,\R)$
(a ``classical Hamiltonian''), and $\SYM{U_{t}}{Ut}:=e^{-\im t\cQ(H)}$,
the corresponding ``quantum time evolution''. (Note that $\widetilde{U_{t}}=e^{-\im t\tilde{\cQ}(H)}E_{\cH^{*}}=E_{\cH^{*}}e^{-\im t\tilde{\cQ}(H)}E_{\cH^{*}}.$)
Then for any $t>0$, the integral kernel of $\widetilde{U_{t}}$ on
$\Manifold\times\Manifold$ is expressed as%
\[
\widetilde{U_{t}}(\p_{1},\p_{2})=\lim_{\no\to\infty}e^{\no tn}\int_{C_{\p_{1},\p_{2}}([0,t],\Manifold)}e^{-\im\int_{0}^{t}H(\path_{s})\d s}\ptrans_{t}^{\path,-1}\d\mu_{0,t;\p_{1},\p_{2}}^{(\no)}(\path),\quad\p_{1},\p_{2}\in\Manifold,
\]
where $n=\dim_{\R}\Manifold$. %
\end{thm}

\end{shaded}
\begin{proof}
Noticing Cor. \ref{thm:siboriQ} %
and
\[
\exp\left[-t\left(\no(\nabla^{*}\nabla-n)+\im H\right)\right]=e^{\no tn}\exp\left[-\no t\left(\nabla^{*}\nabla+\im\no^{-1}H\right)\right],
\]
let $V:=\im\no^{-1}H$. We see
\[
\ptrans_{t}^{\p}(\path^{(\no)})=\ptrans_{\no t}^{\p}(\path),\qquad\int_{0}^{t}H(\path_{s}^{(\no)})\d s=\int_{0}^{\no t}\no^{-1}H(\path_{t})\mathrm{d}s.
\]
Therefore, by (\ref{eq:e^itH(V)(x,y)=00003Ddmu}),%
\begin{align*}
 & \exp\left[-t\left(\no(\nabla^{*}\nabla-n)+\im H\right)\right](\p_{1},\p_{2})\\
 & \quad=e^{\no tn}\exp\left[-\no t\left(\nabla^{*}\nabla+\im\no^{-1}H\right)\right](\p_{1},\p_{2})\\
 & \quad=e^{\no tn}\int_{C_{\p_{1},\p_{2}}([0,\no t],M)}e^{-\int_{0}^{\no t}\im\no^{-1}H(\path_{t})\mathrm{d}s}\ptrans_{\no t}^{\p_{1},-1}(\path)\d\mu_{0,\no t;\p_{1},\p_{2}}(\path)\\
 & \quad=e^{\no tn}\int_{C_{\p_{1},\p_{2}}([0,t],M)}e^{-\im\int_{0}^{t}H(\path_{s}^{(\no)})\d s}\ptrans_{t}^{\p_{1},-1}(\path^{(\no)})\d\mu_{0,t;\p_{1},\p_{2}}^{(\no)}(\path^{(\no)}),
\end{align*}
and hence

\begin{align*}
e^{-\im t\tilde{\cQ}(H)}(\p_{1},\p_{2}) & =\lim_{\no\to\infty}\exp\left[-t\left(\no(\nabla^{*}\nabla-n)+\im H\right)\right](\p_{1},\p_{2})\\
 & =\lim_{\no\to\infty}e^{\no tn}\int_{C_{\p_{1},\p_{2}}([0,t],M)}e^{-\im\int_{0}^{t}H(\path_{s}^{(\no)})\d s}\ptrans_{t}^{\p_{1},-1}(\path^{(\no)})\d\mu_{0,t;\p_{1},\p_{2}}^{(\no)}(\path^{(\no)}).
\end{align*}

\end{proof}
Recall Lemma \ref{thm:kernel} %
and Eq.(\ref{eq:kernel-bundle})%
. Then we can define the integral kernel of $\widetilde{U_{t}}$ on
$\bbS(\Manifold)\times\bbS(\Manifold)$ (not on $\Manifold\times\Manifold$)
by
\[
\SYM{\widetilde{U_{t}}}{Uttil}(v_{1},v_{2}):=\bigl\langle v_{2}|e^{-\im t\cQ(H)}v_{1}\bigr\rangle=\bigl(\widetilde{U_{t}}(\p_{1},\p_{2})v_{2}^{*}\bigr)(v_{1}),\quad v_{k}\in\ran(\p_{k}),\ k=1,2.
\]
Similarly, the parallel transport $\ptrans_{t}^{\path,-1}\in\Hom(\scO_{1,\Manifold,\p_{1}},\scO_{1,\Manifold,\p_{2}})=\Hom(\ran(\p_{1})^{*},\ran(\p_{2})^{*})$
can also viewed as a function on $\ran(\p_{1})\times\ran(\p_{2})$:
\[
\SYM{\ptrans_{t}^{\path,-1}}{//tx}(v_{1},v_{2}):=\left(\ptrans_{t}^{\path,-1}v_{1}^{*}\right)(v_{2}),\quad(v_{1},v_{2})\in\ran(\p_{1})\times\ran(\p_{2}).
\]
If $t_{1}>t_{2}$, let $[t_{1},t_{2}]$ refer to the closed interval
$[t_{2},t_{1}]$, and let $\mu_{t_{2},t_{1};\p_{2},\p_{1}}^{(\no)}$
denote the same measure as $\mu_{t_{1},t_{2};\p_{1},\p_{2}}^{(\no)}$.

\begin{shaded}%
\begin{cor}
\label{cor:<v1|e-itQ(H)v2>=00003D}For any $t_{1},t_{2}\in\R$, $v_{1},v_{2}\in\bbS(\Manifold)$
with $\p_{k}:=\Proj(v_{k})\in\Manifold$ ($k=1,2$),
\begin{align*}
 & \bigl\langle v_{2}|e^{-\im(t_{2}-t_{1})\cQ(H)}v_{1}\bigr\rangle=\widetilde{U_{t_{2}-t_{1}}}(v_{1},v_{2})\\
 & \quad=\lim_{\no\to\infty}e^{n\no|t_{2}-t_{1}|}\int_{C_{\p_{1},\p_{2}}([t_{1},t_{2}],\Manifold)}e^{-\im\int_{t_{1}}^{t_{2}}H(\path_{s})\d s}\ptrans_{t_{2}-t_{1}}^{\path,-1}(v_{1},v_{2})\d\mu_{t_{1},t_{2};\p_{1},\p_{2}}^{(\no)}(\path).
\end{align*}
\end{cor}

\end{shaded}

For $k=1,...,N$, let $t_{1}<\cdots<t_{N}$, $\p_{k}\in\Manifold$,
$v_{k}\in\ran(\p_{k})$, $\left\Vert v_{k}\right\Vert =1$, and $\path_{k}:[t_{k},t_{k+1}]\to\Manifold$
($k=1,...,N-1$) be a path (of a Brownian motion) on $\Manifold$
with $\path_{k}(t_{k})=\p_{k}$, $\path_{k}(t_{k+1})=\p_{k+1}$. Let
$\path:[t_{1},t_{N}]\to\Manifold$ be the concatenation of the paths
$\path_{1},...,\path_{N-1}$. When $N=3$, we have
\[
\ptrans_{t_{N}}^{\path,-1}(v_{1},v_{3})=\left(\ptrans_{t_{3}}^{\path_{2},-1}\ptrans_{t_{2}}^{\path_{1},-1}\right)(v_{1},v_{3})=\left(\ptrans_{t_{3}}^{\path_{2},-1}\ptrans_{t_{2}}^{\path_{1},-1}v_{1}^{*}\right)(v_{3})=\bigl\langle v_{3}^{*}|\ptrans_{t_{3}}^{\path_{2},-1}\ptrans_{t_{2}}^{\path_{1},-1}v_{1}^{*}\bigr\rangle_{\bbS}
\]
\[
=\bigl\langle v_{3}^{*}|\ptrans_{t_{3}}^{\path_{2},-1}v_{2}^{*}\bigr\rangle_{\bbS}\bigl\langle v_{2}^{*}|\ptrans_{t_{2}}^{\path_{1},-1}v_{1}^{*}\bigr\rangle_{\bbS}=\prod_{k=1}^{2}\ptrans_{t_{k+1}}^{\path_{k},-1}(v_{k},v_{k+1}),
\]
and in general
\[
\ptrans_{t_{N}}^{\path,-1}(v_{1},v_{N})=\left(\ptrans_{t_{N}}^{\path_{N-1},-1}\cdots\ptrans_{t_{2}}^{\path_{1},-1}\right)(v_{1},v_{N})=\prod_{k=1}^{N-1}\ptrans_{t_{k+1}}^{\path_{k},-1}(v_{k},v_{k+1})
\]
Let $\vec{\p}=(\p_{1},...,\p_{N})\in\Manifold^{N}$, $\vec{t}=(t_{1},...,t_{N})$,
$0\le t_{1}<\cdots<t_{N}$. 
\[
\SYM{C_{\vec{t},\vec{\p}}}{Ctp}:=\left\{ \path\in C([t_{1},t_{N}],\Manifold)|\path(t_{j})=\p_{j},\ j=1,...,N\right\} 
\]
Consider the measures $\mu_{t_{j},t_{j+1};\p_{j},\p_{j+1}}^{(\no)}$
on $C_{\p_{j},\p_{j+1}}([t_{j},t_{j+1}],\Manifold)$, and define the
measure $\SYM{\mu_{\vec{t},\vec{\p}}^{(\no)}}{mutp(t)}$ on $C_{\vec{t},\vec{\p}}$
by 
\[
\d\mu_{\vec{t},\vec{\p}}^{(\no)}(\path):=\prod_{j=1}^{N-1}\d\mu_{t_{j},t_{j+1};\p_{j},\p_{j+1}}^{(\no)}(\path|_{[t_{j},t_{j+1}]})
\]

\begin{shaded}%
\begin{thm}
\label{thm:prodUk=00003D}For any $0\le t_{1}<\cdots<t_{N}$, $v_{1},...,v_{N}\in\bbS(\Manifold)$
with $\p_{k}:=\Proj(v_{k})\in\Manifold$ ($k=1,...,N$), we have
\[
\prod_{k=1}^{N-1}\left(\widetilde{U}_{t_{k+1}-t_{k}}(v_{k},v_{k+1})\right)=\lim_{\no\to\infty}e^{n\no|t_{N}-t_{1}|}\int_{C_{\vec{t},\vec{\p}}}e^{-\im\int_{t_{1}}^{t_{N}}H(\path_{s})\d s}\ptrans_{t_{N}}^{\path,-1}(v_{1},v_{N})\d\mu_{\vec{t},\vec{\p}}^{(\no)}(\path).
\]
\end{thm}

\end{shaded}
\begin{rem}
We see
\begin{align*}
 & \prod_{k=1}^{N-1}\left(\widetilde{U}_{t_{k+1}-t_{k}}(v_{k},v_{k+1})\right)\\
 & =\bigl\langle v_{N}|e^{-\im(t_{N}-t_{N-1})\cQ(H)}v_{N-1}\bigr\rangle\cdots\bigl\langle v_{2}|e^{-\im(t_{2}-t_{1})\cQ(H)}v_{1}\bigr\rangle\\
 & =\bigl\langle v_{N}|e^{-\im(t_{N}-t_{N-1})\cQ(H)}\p_{N-1}\cdots e^{-\im(t_{3}-t_{2})\cQ(H)}\p_{2}e^{-\im(t_{2}-t_{1})\cQ(H)}v_{1}\bigr\rangle\\
 & =\bigl\langle v_{N}|e^{-\im t_{N}\cQ(H)}\p_{N-1,t_{N-1}}\cdots\p_{2,t_{2}}e^{\im t_{1}\cQ(H)}v_{1}\bigr\rangle
\end{align*}
where $\SYM{\p_{j,t_{j}}}{pjt}:=e^{\im t_{j}\tilde{\cQ}(H)}\p_{j}e^{-\im t_{j}\tilde{\cQ}(H)}.$
Thus, this is invariant under the transformation $v_{k}\mapsto e^{\im\theta_{k}}v_{k}$,
$\theta_{k}\in\R$, for $k=2,...,N-1$. Furthermore, if $v_{1}=v_{N}$,
this equals%
\[
\Tr\left(\p_{N-1,t_{N-1}}\cdots\p_{2,t_{2}}\p_{1,t_{1}}e^{-\im(t_{N}-t_{1})\cQ(H)}\right)
\]
and hence is invariant under the above transformation for all $v_{k}$,
$k=1,...,N$ with $\theta_{1}=\theta_{N}$. 
\end{rem}

\begin{proof}
By Corollary \ref{cor:<v1|e-itQ(H)v2>=00003D}, 
\begin{align*}
 & \prod_{k=1}^{N-1}\widetilde{U}_{t_{k+1}-t_{k}}(v_{k},v_{k+1})\\
 & \quad=\prod_{k=1}^{N-1}\lim_{\no\to\infty}e^{\no(t_{k+1}-t_{k})n}\int_{C_{\p_{j},\p_{j+1}}([t_{k},t_{k+1}],\Manifold)}e^{-\im\int_{t_{k}}^{t_{k+1}}H(\path_{s})\d s}\\
 & \quad\quad\times\ptrans_{t_{k+1}}^{\path_{k},-1}(v_{k},v_{k+1})\d\mu_{t_{k},t_{k+1};\p_{k},\p_{k+1}}^{(\no)}(\path_{k})\\
 & \quad=\lim_{\no\to\infty}e^{\no(t_{N}-t_{1})n}\left(\prod_{k=1}^{N-1}\int_{C_{\p_{j},\p_{j+1}}([t_{k},t_{k+1}],\Manifold)}\d\mu_{t_{k},t_{k+1};\p_{k},\p_{k+1}}^{(\no)}(\path_{k})\right)\\
 & \quad\quad\times\prod_{k=1}^{N-1}e^{-\im\int_{t_{k}}^{t_{k+1}}H(\path_{s})\d s}\ptrans_{t_{k+1}}^{\path_{k},-1}(v_{k},v_{k+1})\\
 & \quad=\lim_{\no\to\infty}e^{\no(t_{N}-t_{1})n}\int_{C_{\vec{t},\vec{\p}}}e^{-\im\int_{t_{1}}^{t_{N}}H(\path_{s})\d s}\ptrans_{t_{N}}^{\path,-1}(v_{1},v_{N})\d\mu_{\vec{t},\vec{\p}}^{(\no)}(\path),
\end{align*}
\end{proof}

\section{Representation of quantum probability}

\label{sec:Representation-of-QP}

Recall (\ref{eq:TrQt1QtN}), then we want a path integral formula
for $\Tr\left(\p_{1,t_{1}}\cdots\p_{N,t_{N}}\right)$ to represent
the trace of product of some operators, which often has a physical
meaning as a quantum probability or an expectation value of some physical
quantity. Roughly speaking, if we set $t_{1}=t_{N}$ in Theorem \ref{thm:prodUk=00003D},
then we could get it. However, Theorem \ref{thm:prodUk=00003D} is
meaningful only when $0\le t_{1}<\cdots<t_{N}$. If we want to extend
Theorem \ref{thm:prodUk=00003D} for any $t_{1},...,t_{N}\in\R$,
we need to introduce a curve parameter $\ppa$ other than the time
parameter $t$. 

Assume that $H:\Manifold\to\R$ is bounded and smooth. Consider the
manifold $\Manifold\times\R$ with the natural Riemannian metric.
Then $\scO_{1,\Manifold}\times\R$ is a line bundle of $\Manifold\times\R$.
Define the connection $\nabla^{(H)}$ on $\scO_{1,\Manifold}\times\R$
by 
\[
\SYM{\nabla_{X\oplus\di_{t}}^{(H)}}{nablaH}s:=\nabla_{X}s+\im H\di_{t}s,\qquad X:\text{ a vector field on }\Manifold,\ s:\text{ a section of }\scO_{1,\Manifold}\times\R,\ \di_{t}:=\frac{\di}{\di t},
\]
where $\nabla$ is the Chern connection on $\Manifold$. For a path
$\Path$ on $\Manifold\times\R$, let $\SYM{\ptrans^{(H)}}{//(H)}(\Path)$
or $\ptrans^{(H),\Path}$ denote the parallel transport along $\Path$
w.r.t. $\nabla^{(H)}$.

Let $t_{0}\le t_{1}$. For $\path\in C([t_{0},t_{1}],\Manifold)$,
define $\tilde{\path},\tilde{\path}^{\inv}\in C([0,1],\Manifold\times\R)$
by
\[
\SYM{\tilde{\path}(\ppa)}{xtil}:=\left(\path\left(t\right),t\right),\quad t:=\ppa\left(t_{1}-t_{0}\right)+t_{0},\ \ppa\in[0,1]
\]
\[
\SYM{\tilde{\path}^{\inv}(\ppa)}{xtil-1}:=\tilde{\path}(1-\ppa),\quad\ppa\in[0,1].
\]
Notice that $\tilde{\path}^{\inv}=(\tilde{\path})^{\inv}$ differs
from $\widetilde{(\path^{\inv})}$. If $\path$ is piecewise smooth,
we find
\begin{equation}
\ptrans_{t}^{(H),\tilde{\path},-1}=e^{-\im\int_{0}^{t}H(\path_{s})\d s}\ptrans_{t}^{\path,-1}.\label{eq://=00003Dexp//}
\end{equation}
When $\path$ is a Brownian motion, this equation can be understood
as a stochastic one.

For $\p_{0},\p_{1}\in\Manifold$ and $t_{0},t_{1}\in\R$, let

\[
\SYM{\tilde{C}_{t_{0},t_{1};\p_{0},\p_{1}}}{Ctilpptt}:=\begin{cases}
\left\{ \tilde{\path}|\ \path\in C_{\p_{0},\p_{1}}([t_{0},t_{1}],\Manifold)\right\}  & \text{if }t_{0}\le t_{1}\\
\left\{ \tilde{\path}^{\inv}|\ \path\in C_{\p_{1},\p_{0}}([t_{1},t_{0}],\Manifold)\right\}  & \text{if }t_{0}>t_{1}
\end{cases}
\]
so that if $\Path_{1}\in\tilde{C}_{t_{0},t_{1};\p_{0},\p_{1}}$ and
$\Path_{2}\in\tilde{C}_{t_{1},t_{2};\p_{1},\p_{2}}$, then the concatenation
$\Path_{2}\bullet\Path_{1}\in C([0,1],\Manifold\times\R)$ is defined:
\[
\left(\Path_{2}\bullet\Path_{1}\right)(\ppa):=\begin{cases}
\Path_{1}(2\ppa) & \left(0\le\ppa\le\frac{1}{2}\right)\\
\Path_{2}(2\ppa-1) & \left(\frac{1}{2}<\ppa\le1\right)
\end{cases}
\]

If $t_{0}\le t_{1}$ (resp. $t_{0}>t_{1}$), the measure $\mu_{t_{0},t_{1};\p_{1},\p_{2}}^{(\no)}$
on $C_{\p_{0},\p_{1}}([t_{0},t_{1}],\Manifold)$ (resp. $C_{\p_{1},\p_{0}}([t_{1},t_{0}],\Manifold)$)
induces the measure $\SYM{\tilde{\mu}_{t_{0},t_{1};\p_{0},\p_{1}}^{(\no)}}{mutil}$
on $\tilde{C}_{t_{0},t_{1};\p_{0},\p_{1}}$. Let $\vec{\p}=(\p_{1},...,\p_{N})\in\Manifold^{N}$,
$\vec{t}=(t_{1},...,t_{N})\in\R^{N}$. Define the space of loops ${\rm Loop}_{\vec{t},\vec{\p}}\subset C([0,1],\Manifold\times\R)$
by
\[
\SYM{{\rm Loop}_{\vec{t},\vec{\p}}}{Loop}:=\left\{ \Path_{N}\bullet\cdots\bullet\Path_{1}|\Path_{k}\in\tilde{C}_{\p_{k},\p_{k+1};t_{k},t_{k+1}},\ k=1,...,N,\ \p_{N+1}:=\p_{1},\ t_{N+1}:=t_{1}\right\} 
\]
where $\Path_{N}\bullet\cdots\bullet\Path_{1}:=\Path_{N}\bullet\left(\Path_{N-1}\bullet\left(\cdots\bullet\Path_{1}\right)\cdots\right)$,
although the order of the concatenations is not important in the following;
That is, we could alternatively define $\Path_{N}\bullet\cdots\bullet\Path_{1}$
by, say, $\left(\cdots\left(\Path_{N}\bullet\Path_{N-1}\right)\bullet\cdots\bullet\Path_{1}\right)$.

Define the measure $\SYM{\mu_{\vec{t},\vec{\p}}^{(\no)}}{mupt}$ on
${\rm Loop}_{\vec{\p},\vec{t}}$ by

\[
\d\mu_{\vec{t},\vec{\p}}^{(\no)}(\Path):=\prod_{k=1}^{N}\d\tilde{\mu}_{t_{k},t_{k+1};\p_{k},\p_{k+1}}^{(\no)}(\Path_{k}),\quad\Path=\Path_{N}\bullet\cdots\bullet\Path_{1}\in{\rm Loop}_{\vec{\p},\vec{t}},
\]
where $\p_{N+1}:=\p_{1},\ t_{N+1}:=t_{1}$. For $\Path\in{\rm Loop}_{\vec{\p},\vec{t}}$,
of course, $\ptrans^{(H)}(\Path)\in\rU(1)$ refers to the (stochastic)
holonomy along the loop $\Path$.

\noindent\begin{minipage}[t]{1\columnwidth}%
\begin{shaded}%
\begin{lem}
\label{thm:<v2|ev1>-loop}For any $t_{1},t_{2}\in\R$, $v_{1},v_{2}\in\bbS(\Manifold)$
with $\p_{k}:=\Proj(v_{k})\in\Manifold$, $k=1,2$,
\[
\bigl\langle v_{2}|e^{-\im(t_{2}-t_{1})\cQ(H)}v_{1}\bigr\rangle=\lim_{\no\to\infty}e^{n\no|t_{2}-t_{1}|}\int_{\tilde{C}_{\p_{1},\p_{2}}([t_{1},t_{2}],\Manifold)}\ptrans_{t_{2}-t_{1}}^{(H),\Path,-1}(v_{1},v_{2})\d\mu_{t_{1},t_{2};\p_{1},\p_{2}}^{(\no)}(\Path).
\]
\end{lem}

\end{shaded}%
\end{minipage}
\begin{proof}
Directly follows from Corollary \ref{cor:<v1|e-itQ(H)v2>=00003D}
and (\ref{eq://=00003Dexp//}).
\end{proof}
\begin{shaded}%
\begin{thm}
\label{thm:Tr(pp)=00003D}Let $\vec{\p}=(\p_{1},...,\p_{N})\in\Manifold^{N}$,
$\vec{t}=(t_{1},...,t_{N})\in\R^{N}$. Then%
\begin{equation}
\Tr\left(\p_{1,t_{1}}\cdots\p_{N,t_{N}}\right)=\lim_{\no\to\infty}e^{nT\no}\int_{{\rm Loop}_{\vec{\p},\vec{t}}}\ptrans^{(H)}(\Path)\d\mu_{\vec{\p},\vec{t}}^{(\no)}(\Path)\label{eq:Tr(pp)=00003D}
\end{equation}
where $T:=\sum_{k=1}^{N}\left|t_{k+1}-t_{k}\right|$ ($t_{N+1}:=t_{1}$),
$\p_{1},...,\p_{N}\in\Manifold$ ($k=1,...,N$).
\end{thm}

\end{shaded}
\begin{proof}
Let $v_{1},...,v_{N}\in\bbS(\Manifold)$ with $\p_{k}=\Proj(v_{k})\in\Manifold$.
Then we see%
\[
\ol{\Tr\left(\p_{1,t_{1}}\cdots\p_{N,t_{N}}\right)}=\Tr\left(\p_{N,t_{N}}\cdots\p_{1,t_{1}}\right)=\prod_{j=1}^{N}\bigl\langle v_{j+1}|e^{-\im(t_{j+1}-t_{j})\cQ(H)}v_{j}\bigr\rangle
\]
where $v_{N+1}:=v_{1}$, $t_{N+1}:=t_{1}$. Hence (\ref{eq:Tr(pp)=00003D})
follows from Lemma \ref{thm:<v2|ev1>-loop}.
\end{proof}
Let $\cB(\Manifold)$ denote the family of Borel sets of $\Manifold$.
Let
\begin{align*}
\SYM{E_{0}(S)}{E0} & :=\cQ(\chi_{S})=\int_{S}\p\,\d\mu(\p),\qquad S\in\cB(\Manifold).\\
\SYM{E_{t}(S)}{Et(S)} & :=e^{-\im t\cQ(H)}E_{0}(S)e^{\im t\cQ(H)}=\int_{S}\p_{t}\,\d\mu(\p),\quad t\in\R.
\end{align*}
For each $t\in\R$, $E_{t}(\bullet)$ is called a \emph{positive operator
valued measure} (POVM) on $\Manifold$. Let $\rho$ be a density operator
on $\cH$, i.e., $\rho\ge0$, $\Tr\rho=1$, and let $0\le t_{1}\le\cdots\le t_{N}$.
Then the value 
\[
\SYM{P_{\rho}(\vec{t},\vec{S})}{Prho}:=\Tr E_{t_{N}}(S_{N})\cdots E_{t_{1}}(S_{1})\rho E_{t_{1}}(S_{1})\cdots E_{t_{N}}(S_{N}),\quad\vec{S}:=(S_{1},...,S_{N})
\]
is interpreted as the joint probability that under the condition that
the state at time 0 is $\rho$, the position in the phase space $\Manifold$
is measured to be in $S_{1}$ at time $t_{1}$, and then the position
in the phase space $\Manifold$ is measured to be in $S_{2}$ at time
$t_{2}$, etc. Of course, these measurements are somewhat ``fuzzy''
in that even if $S,S'\in\cB(\Manifold)$ satisfy $S\cap S'=\emptyset$,
the probability $P_{\rho}((t,t),(S,S'))$ can be non-zero; Any error-free
quantum measurement on $\Manifold$ is impossible by the uncertainty
principle.

\begin{shaded}%
\begin{cor}
\label{thm:jointProb}Let $\rho$ be a density operator which have
the representation $\rho=\cQ(f_{\rho})$ for some Borel function $f_{\rho}:\Manifold\to\R$.
Let
\begin{align*}
(F_{1},...,F_{2N+1}) & :=(\chi_{S_{N}},...,\chi_{S_{1}},f_{\rho},\chi_{S_{1}},...,\chi_{S_{N}}),\\
(\tau_{1},...,\tau_{2N+1}) & :=(t_{N},...,t_{1},t_{0},t_{1},...,t_{N}),\qquad t_{0}:=0.
\end{align*}
Then we have the path-integral representation of the quantum probability
\[
P_{\rho}(\vec{t},\vec{S})=\lim_{\no\to\infty}e^{2nt_{N}\no}\int_{\Manifold^{2N+1}}\d\mu^{2N+1}(\vec{\p})\int_{{\rm Loop}_{\vec{\p},\vec{\tau}}}\d\mu_{\vec{\p},\vec{t}}^{(\no)}(\Path)\ptrans^{(H)}(\Path)\prod_{j=1}^{2N+1}F_{j}(\p_{j}),
\]
where $\vec{\p}=(\p_{1},...,\p_{2N+1})$, $\vec{\tau}=(\tau_{1},...,\tau_{2N+1})$,
and
\[
\int_{\Manifold^{2N+1}}\d\mu^{2N+1}(\vec{\p})\ \text{ denotes }\int_{\Manifold}\d\mu(\p_{1})\cdots\int_{\Manifold}\d\mu(\p_{2N+1}).
\]
\end{cor}

\end{shaded}
\begin{proof}
By Theorem \ref{thm:Tr(pp)=00003D} and Eq.(\ref{eq:TrQt1QtN}), we
find%
\begin{align*}
P_{\rho}(\vec{t},\vec{S}) & =\Tr\cQ_{t_{N}}(\chi_{S_{N}})\cdots\cQ_{t_{1}}(\chi_{S_{1}})\cQ(f_{\rho})\cQ_{t_{1}}(\chi_{S_{1}})\cdots\cQ_{t_{N}}(\chi_{S_{N}})\\
 & =\int_{\Manifold}\d\mu(\p_{1})\cdots\int_{\Manifold}\d\mu(\p_{2N+1})\Tr\left(\p_{1,t_{1}}\cdots\p_{N,t_{N}}\right)\prod_{j=1}^{N}F_{j}(\p_{j})\\
 & =\int_{\Manifold}\d\mu(\p_{1})\cdots\int_{\Manifold}\d\mu(\p_{2N+1})\left[\prod_{j=1}^{2N+1}F_{j}(\p_{j})\right]\lim_{\no\to\infty}e^{nT\no}\int_{{\rm Loop}_{\vec{\p},\vec{t}}}\ptrans^{(H)}(\Path)\d\mu_{\vec{\p},\vec{t}}^{(\no)}(\Path),
\end{align*}
with
\[
T:=\sum_{k=1}^{2N}|\tau_{k+1}-\tau_{k}|=2\sum_{j=0}^{N-1}(t_{k+1}-t_{k})=2t_{N}.
\]
\end{proof}
\begin{rem}
Note that there exist many density operators $\rho$ which do not
have the representation $\rho=\cQ(f_{\rho})$. However, the set of
density operators of the form $\cQ(f)$ is dense in the space of density
operators, which has a metric induced by the trace norm, and hence
any density operator can be approximated by the operator of the form
$\cQ(f)$.
\end{rem}

\begin{rem}
We also emphasize that the formulas for quantum joint probabilities
as Corollary \ref{thm:jointProb} will not be formulated in the \emph{imaginary-time}
path integral; There will be no \emph{direct} relation between the
quantum joint probability (with real-time evolution) and the imaginary-time
path integral.
\end{rem}

\section{Example: Glauber coherent states}

Consider the case where $\Manifold\subset\bbP\cH$ is homeomorphic
to $\C^{n}\cong\R^{2n}$. Let $\psi:\C^{n}\to\cH^{\times}$ be a holomorphism
such that $\Proj\circ\psi(\C^{n})=\Manifold$, so that $\Proj\circ\psi:\C^{n}\to\Manifold$
is a single global coordinate chart of $\Manifold$. Without loss
of generality, assume $\left\langle \psi(0)|\psi(z)\right\rangle =1$
for all $z\in\C^{n}$.

The line bundle $\scO_{1,\Manifold}$ over $\Manifold$, which is
topologically trivial, i.e.~$\cong\Manifold\times\C$, has the Hermitian
metric $\left\langle \cdot|\cdot\right\rangle _{{\rm pt}}$ given
by $\left\langle f_{1}|f_{2}\right\rangle _{{\rm pt}}(\Proj(v)):=\left\Vert v\right\Vert ^{-2}\ol{f_{1}\left(v\right)}f_{2}\left(v\right)$,
where $v\in\Proj^{-1}(\Manifold),\ f_{1},f_{2}\in\CC_{1,\Manifold}.$
A global K\"ahler potential $\Kpot:\Manifold\to\R$ is given by

\[
\Kpot(z)=\frac{1}{h(z)}=\left\Vert \psi(z)\right\Vert ^{2}.
\]
The corresponding global symplectic form is $\omega:=-\im\,\d''\d'\log\Kpot$,
and the Riemannian metric is $g(X,Y):=\omega(X,JY)$. Let $\ebold:=\psi(0)$
then $\ebold^{*}$ is a global holomorphic section of $\scO_{1,\Manifold}$;
Precisely, if we identify the fiber of $\scO_{1,\Manifold}$ over
$\p\in\Manifold$ with $\ran(\p)^{*}$, the value $\ebold^{*}(\p)$
of the section $\ebold^{*}\in\CC_{1,\Manifold}$ at $\p\in\Manifold$
is the function $\ran(\p)\to\C$, $v\mapsto\left\langle \ebold|v\right\rangle $.
The normalization of $\ebold^{*}$ is $\ul{\ebold}^{*}:=h^{-1/2}\ebold^{*}$;
Precisely, the value $\ul{\ebold}^{*}(\p)$ of the section $\ul{\ebold}^{*}\in\CC_{1,\Manifold}$
at $\p\in\Manifold$ is the function
\[
\ran(\p)\to\C,\qquad\zeta\ul{\psi(z)}\mapsto\zeta,\qquad z\in\C^{n},\ \psi(z)\in\ran(\p),\ \zeta\in\C.
\]
Here the Chern connection form w.r.t.~the holomorphic frame $\ebold^{*}$
is a $\C$-valued 1-form on $\C^{n}$ defined by $\theta:=\d'\log h=-\d'\log\Kpot_{\io}$.
The Chern connection is globally defined by
\[
\nabla\left(f\ebold^{*}\right):=\left(\d f\right)\otimes\ebold^{*}+f\theta\otimes\ebold^{*},\qquad f\in\CC_{0,\Manifold}^{\infty}.
\]
The Chern connection form $\theta_{{\rm nor}}$ on $\C^{n}$ w.r.t.~the
normalized frame $\ul{\ebold}^{*}$ is defined by (\ref{eq:def:thetanor}).%

Let $C:[0,1]\to\C^{n}$ be a piecewise smooth curve, so that $\path:=\Proj\circ\psi\circ C$
is a piecewise smooth curve on $\Manifold$. %
{} If $C$ is a loop, i.e.~$C(0)=C(1)$, the parallel transport (holonomy)
$\ptrans_{1}^{\path}$ becomes a scalar:
\[
\ptrans^{\path}\equiv\ptrans_{1}^{\path}=e^{\im\int_{C}\theta_{{\rm nor}}}\in\rU(1).
\]

Assume the conditions of Corollary \ref{thm:jointProb}. Let $\Path\in{\rm Loop}_{\vec{t},\vec{\p}}$,
with $\Path(\ppa)=(\path(\ppa),\gamma(\ppa))$ ($\ppa\in[0,1]$) where
$\path:[0,1]\to\Manifold$, $\gamma:[0,1]\to\R$. Consider the manifold
$\Manifold\times\R$ as in Sec.~\ref{sec:Representation-of-QP},
and let $H:\Manifold\to\R$ be a bounded smooth Hamiltonian. The connection
on the line bundle $\scO_{1,\Manifold}\times\R$ over $\Manifold\times\R$
defined in Sec.~\ref{sec:Representation-of-QP} is explicitly given
by the global $\R$-valued 1-form
\[
\Theta_{H}:=\theta_{{\rm nor}}+H\d t,
\]
on $\C^{n}\times\R$, w.r.t.~the normalized frame of $\scO_{1,\Manifold}\times\R$
, and hence the holonomy along $\Path$ becomes
\[
\ptrans^{(H)}(\Path)=\exp\left(\im\int_{\hat{\Path}}\Theta_{H}\right),\qquad\hat{\Path}(\ppa):=(\psi^{-1}(\path(\ppa)),\gamma(\ppa))\in\C^{n}\times\R.
\]
Let $\widehat{{\rm Loop}}_{\vec{t},\vec{\p}}:=\left\{ \hat{\Path}|\Path\in{\rm Loop}_{\vec{t},\vec{\p}}\right\} $.
The measure $\mu_{\vec{\p},\vec{t}}^{(\no)}$ on ${\rm Loop}_{\vec{t},\vec{\p}}$
induces the measure $\hat{\mu}_{\vec{\p},\vec{t}}^{(\no)}$ on $\widehat{{\rm Loop}}_{\vec{t},\vec{\p}}$
by the bijection $\Path\mapsto\hat{\Path}$. Now the quantum joint
probability formula in Corollary \ref{thm:jointProb} is rewritten
in a somewhat more familiar and intuitive form:

\begin{shaded}%
\begin{thm}
\label{thm:jointProbCn}Under the conditions of Corollary \ref{thm:jointProb},%
\[
P_{\rho}(\vec{t},\vec{S})=\lim_{\no\to\infty}e^{2nt_{N}\no}\int_{\Manifold^{2N+1}}\d\mu^{2N+1}(\vec{\p})\int_{\widehat{{\rm Loop}}_{\vec{\p},\vec{\tau}}}\d\hat{\mu}_{\vec{\p},\vec{t}}^{(\no)}(\hat{\Path})e^{\im\int_{\hat{\Path}}\Theta_{H}}\prod_{j=1}^{2N+1}F_{j}(\p_{j}).
\]
\end{thm}

\end{shaded}
\begin{example}
Next let us consider the most basic but important case. Let $n=1$.
Let $a^{*},a$ be usual creation/annihilation operators on a Hilbert
space $\cH$, which satisfy $[a,a^{*}]=1$, and assume that $\{a,a^{*}\}$
is irreducible on $\cH$. Let $v_{0}\in\bbS(\cH)$ be a ``vacuum
state, '' i.e.~$av_{0}=0$. Set
\[
\psi(z)=e^{za^{*}}v_{0},\qquad z\in\C
\]
then the corresponding K\"ahler potential, the symplectic form, and
the Riemannian metric are calculated as
\[
\Kpot(z)=\left\Vert \psi(z)\right\Vert ^{2}=e^{|z|^{2}},
\]
\[
\omega=-\im\,\d''\d'\log\Kpot=\im\,\d z\wedge\d\ol z=2\d x\wedge\d y,\qquad g=2\d x\otimes\d y.
\]
Thus $\Manifold\cong\C$ as K\"ahler manifolds; The volume measure
on $\Manifold$ is the usual Lebesgue measure $\d x\d y$ on $\C\cong\R^{2}$
times 2. %
Note that the normalized vector $\ket z:=\ul{\psi(z)}:=e^{-|z|^{2}/2}\psi(z)$
is nothing other than the coherent state in the usual sense (i.e.~the
\emph{Glauber coherent state}). We can also check $\left\langle \psi(0)|\psi(z)\right\rangle =1$.
The overcompleteness relation 
\[
\frac{1}{\pi}\int_{\C}\ket z\bra z\d x\d y=I,\qquad z=x+\im y
\]
implies that we should take the measure $\mu$ on $\Manifold$ as
$\d\mu(z):=\frac{1}{\pi}\d x\d y$. Now the BS quantization of $f:\Manifold\to\R$
is given by
\[
\cQ(f)=\int_{\C}\hat{f}(z)\ket z\bra z\d\mu(z),\quad\text{where }\hat{f}(z):=f(\Proj(\psi(z)))=f(\ket z\bra z).
\]
The r.h.s.~is often called the Glauber--Sudarshan representation
of the operator of l.h.s.

$\ebold^{*}=\psi(0)^{*}=v_{0}^{*}$ is a holomorphic section of $\scO_{1,\Manifold}$.
By (\ref{eq:def:thetanor}) we find
\[
\theta_{{\rm nor}}=y\d x-x\d y,\qquad z=x+\im y.
\]
Let $C:[0,1]\to\C$ be a piecewise smooth curve, with $\path:=\Proj\circ\psi\circ C$.
Then we find that the operation of the parallel transport $\ptrans_{t}^{\path}$
on $\scO_{1,\Manifold}$ is explicitly written as
\[
\ptrans_{t}^{\path}\bra{C(0)}=e^{\im\int_{C\upharpoonright[0,t]}\theta_{{\rm nor}}}\bra{C(t)},
\]
where we used the bra notation $\bra z:=\ul{\psi(z)^{*}}$, $z\in\C$.%
{} Let $\Theta_{H}:=\theta_{{\rm nor}}+H\d t$, then the quantum joint
probability w.r.t.~the time evolution generated by the quantum Hamiltonian
$\cQ(H)$ is given by Theorem \ref{thm:jointProbCn}.

Note that the paths occurring in the (stochastic) line integral in
this formula are closed, and hence even if we substitute an arbitrary
$\theta_{{\rm nor}}':=\theta_{{\rm nor}}+\alpha$ with $\d\alpha=0$
for $\theta_{{\rm nor}}$, we get the same probability. To fix the
1-form $\theta_{{\rm nor}}$ is analogous to a gauge fixing in physics;
This suggests that Theorem \ref{thm:jointProbCn} realizes a somewhat
``gauge-invariant'' formulation of quantization. However, of course,
different gauge fixings of a classical system can lead to different
quantizations, where the difference is experimentally observable,
in general. Thus we can only say that Theorem \ref{thm:jointProbCn}
may \emph{reduce} the ``gauge-dependence'' of the notion of quantizations.
\end{example}

\section{Toward the geometric path integral quantizations}

The results in the previous sections are given in the situation where
the BS quantization $\cQ(f)$ is considered only when the function
$f$ is bounded, and so $\cQ(f)$ becomes a bounded operator. The
various difficulties in dealing with unbounded operators put obstacles
in the generalizations of those results, and so it may be very difficult
to formulate a general theory. (One can see such difficulty also from
the case of \emph{deformation quantization}; The theory of deformation
quantization was presented by \cite{BFFLS1} in 1978. The theory achieved
a great success in the \emph{algebraic level}, when Kontsevich \cite{Kon97}
proved its generality, that there exists a deformation quantization
for every Poisson manifold. However, the theories to realize the deformation
quantizations in terms of the operators in the Hilbert spaces (e.g.
the \emph{strict deformation quantization} of Rieffel \cite{Rie89,Rie93})
seem to remain incomplete even on the problems concerning only bounded
operators.) 

Nevertheless, we conjecture that the results of the previous sections
can be generalized for most situations which are ``physically relevant.''
To clarify this conjecture, we consider the notion of ``geometric
path integral quantization'' in this section.

Let $\Manifold$ be a complete K\"ahler manifold, and assume the
condition of Proposition \ref{thm:bridge} as a Riemannian manifold.
(Forget the assumption in Sec.~\ref{sec:BSquantization} that $\Manifold$
is a submanifold of some projective space $\bbP\cH$.) Physically,
we interpret $\Manifold$ as a classical-mechanical phase space, whose
symplectic form is the K\"ahler form $\omega$. Recall the notion
of prequantization bundle in \emph{geometric quantization} \cite{Woo92}.
\begin{defn}
A symplectic manifold $(\Manifold,\omega)$ is \termi{prequantizable}
when there exists a Hermitian line bundle, called a \termi{prequantization bundle},
$\pi:L\rightarrow\Manifold$ with connection $\nabla$, whose curvature
form $\Theta$ is proportional to the symplectic 2-form, %
$\Theta=-\im\omega/\hbar$. (Cf. Eq.(\ref{eq:omega=00003DiTheta}))

Note: In this paper, we set $\hbar=1$, and recall that $\Theta$
is defined to be a $\fraku(1)=\im\R$-valued 2-form here.
\end{defn}

For quantization, we must assume that $\Manifold$ is prequantizable.
Moreover we assume that the prequantization bundle $L\to\Manifold$
is holomorphic. 

\begin{shaded}%
\begin{conjecture}
\label{conj:ext0}Assume that $H:\Manifold\to\R$ satisfy some adequate
conditions as a classical-mechanical Hamiltonian. Then Cor. \ref{thm:jointProb}
can be generalized for such $H$.%
\end{conjecture}

\end{shaded}
\begin{rem}
In the typical cases, $H(x)$ is smooth, bounded from below, and increases
in a polynomial order as $\left\Vert x\right\Vert \to\infty$. Hence
we could take the above ``adequate conditions'' as such conditions.
However also note the important cases such as the Hamiltonian of a
hydrogen atom, which is not bounded from below.
\end{rem}

Note that since a Brownian motion on $\Manifold$ is well-defined,
our path integral is also well-defined for each fixed $\no$. The
above assertion says that the probability $P_{\rho}(\vec{t},\vec{S})$
calculated via our path integral (with the limit $\no\to\infty$)
coincides with the value calculated via BS quantization. However,
since $\Manifold$ is not assumed to be a subset of $\bbP\cH$ here,
BS quantization is not defined yet, and so the above assertion is
still too vague. %
{} We will explain further this point in the following.

Consider the Hilbert space $\cK:=\Gamma_{{\rm hol}}^{L^{2}}\subset\Gamma^{L^{2}}$,
where $\Gamma^{L^{2}}$ is the space of $L^{2}$ sections of $B$,
and $\Gamma_{{\rm hol}}^{L^{2}}$ is the closed subspace consisting
of holomorphic sections. Here we consider $\cK$ as the quantum state
space, following the method of \termi{holomorphic quantization}.

Let $K_{A}$ be the integral kernel of an operator $A$ on $\Gamma^{L^{2}}$,
i.e., 
\[
\left(As\right)(x_{1})=\int_{\Manifold}K_{A}(x_{1},x_{2})s(x_{2})\,\d x_{2},\quad s\in\Gamma^{L^{2}},\ x_{1}\in\Manifold.
\]
where $\d x_{2}$ denotes the integral w.r.t. the volume form $\vol$
on $\Manifold$. $K_{A}$ is a map such that $K_{A}(x_{1},x_{2})\in\Hom(L_{x_{2}},L_{x_{1}})$
for all $x_{1},x_{2}\in\Manifold$, where $L_{x}$ is the fiber of
the line bundle $L$ at $x\in\Manifold$; Equivalently, $K_{A}$ is
a section of the external tensor product bundle $L\boxtimes L^{*}\to\Manifold\times\Manifold$. 

Let $E_{\cK}$ denote the orthogonal projection from $\Gamma^{L^{2}}$
onto $\cK$. For each $x\in\Manifold$, define $v_{x}\in\bbS(\cK)\subset\Gamma^{L^{2}}$
by 
\[
v_{x}(x'):=C_{x}K_{E_{\cK}}(x',x),\quad x,x'\in\Manifold,\ C_{x}>0.
\]
Define $P:\Manifold\to\bbP\cK$ by
\[
P(x):=\ket{v_{x}}\bra{v_{x}},\qquad x\in\Manifold.
\]
Assume that this map $P$ is an embedding of the K\"ahler manifold
$\Manifold$ into $\bbP\cK$, viewed as a possibly infinite-dimensional
K\"ahler manifold. This assumption says that $\Manifold$ may be
identified with its range $P(\Manifold)\subset\bbP\cK$, in other
words, that $\Manifold$ can be seen as a submanifold of the projective
space $\bbP\cK$. In this situation the BS quantization can be defined
for $\Manifold$, and so the meaning of Conjecture \ref{conj:ext0}
becomes clearer.
\begin{acknowledgement*}
I am grateful to Koichi Arashi for several useful remarks on a preliminary
version of the manuscript.
\end{acknowledgement*}

\phantomsection	
\addcontentsline{toc}{section}{Reference}



\providecommand{\noopsort}[1]{}\providecommand{\singleletter}[1]{#1}%

\end{document}